\newif\ifreport\reporttrue
\newif\ifblinded\blindedfalse
\documentclass[10pt, conference]{IEEEtran}

\usepackage[usenames]{color}
\usepackage{amsmath, amssymb, xspace, epsfig, graphicx, url, tabularx}
\usepackage{setspace,comment}
\usepackage{mathpartir}
\usepackage{wrapfig}
\usepackage{subfigure}
\usepackage{psfrag}
\usepackage{utf8,times}
\usepackage{paralist}
\usepackage{multirow}
\usepackage{listings}
\usepackage[table,xcdraw]{xcolor}
\usepackage{declarations}
\usepackage{ stmaryrd }
\usepackage{balance}
\usepackage{esvect}
\usepackage{tikz}
\usetikzlibrary{automata,positioning}
\usepackage{amsfonts}

\usepackage{enumitem}
\usepackage[ruled]{algorithm}
\usepackage{algpseudocode}

\usepackage{pdflscape}

\usepackage{wasysym}

\usepackage{array}
\newcommand{\PreserveBackslash}[1]{\let\temp=\\#1\let\\=\temp}
\newcolumntype{C}[1]{>{\PreserveBackslash\centering}p{#1}}
\newcolumntype{R}[1]{>{\PreserveBackslash\raggedleft}p{#1}}
\newcolumntype{L}[1]{>{\PreserveBackslash\raggedright}p{#1}}

\usepackage[colorlinks]{hyperref}

\usepackage{soul}
\newcommand{\hout}[1]{\ifmmode\text{\sout{\ensuremath{#1}}}\else\sout{#1}\fi}

\newif\ifdelimited\delimitedfalse
\newif\ifstaticni\staticnifalse
\newif\iftightgr\tightgrtrue

\title{Towards a General-Purpose Dynamic Information Flow Policy}

\ifblinded
\author{Anonymized for Review}{}{}
\else
\author{\IEEEauthorblockN{Peixuan Li, Danfeng Zhang}
\IEEEauthorblockA{Department of Computer Science and Engineering \\ 
Pennsylvania State University, University Park, PA  United States \\
e-mail: \{pzl129,zhang\}@cse.psu.edu}}  
\fi

\begin{document}

\maketitle
\thispagestyle{plain}
\pagestyle{plain}
\begin{abstract}
Noninterference offers a rigorous end-to-end guarantee for secure
propagation of information. However, real-world systems almost always involve
security requirements that change during program execution, making
noninterference inapplicable. Prior works alleviate the limitation to some
extent, but even for a veteran in information flow security, understanding the
subtleties in the syntax and semantics of each policy is challenging, largely
due to very different policy specification languages, and more fundamentally,
semantic requirements of each policy. 

We take a top-down approach and present a novel information flow
policy, called \policyname{}, which allows information flow restrictions to
downgrade and upgrade in arbitrary ways. \policyname{} is formalized on a novel
framework that, for the first time, allows us to compare and contrast various
dynamic policies in the literature. We show that \policyname{} generalizes
declassification, erasure, delegation and revocation. Moreover, it is
the only dynamic policy that is both applicable and correct on a benchmark of
tests with dynamic policy.
\end{abstract}


\newif\ifignore\ignorefalse
\newif\ifforgetful\forgetfultrue
\newif\ifscopedsecrets\scopedsecretsfalse

\section{Introduction}
\label{sec:introduction}

While noninterference~\cite{noninterference} has become a clich\'{e} for
end-to-end data confidentiality and integrity in information flow security,
this well-accepted concept only describes the ideal security expectations in a
\emph{static} setting, i.e., when data sensitivity \emph{does not} change
throughout program execution. However, real-world applications almost always
involve some dynamic security requirements, which motivates the development of
various kinds of \emph{dynamic} information flow policies:
\begin{itemize}
\item A \emph{declassification
policy}~\cite{askarov2007,flowspecs,cohen1978,sabelfeld2001,
giacobazzi2004,giacobazzi2005,li2005,sabelfeld2003} weakens noninterference by
deliberately releasing (i.e., declassifying) sensitive information. For instance, a
conference management system typically allows deliberate release of paper
reviews and acceptance/rejection decisions after the notification time.

\item An \emph{erasure policy}~\cite{chong2005, chong2008, hunt2008, del2011,
hansen2006, askarov2015} strengthens noninterference by requiring some public
information to become more sensitive, or be erased completely when certain
condition holds.  For example, a payment system should not retain any record of
credit card details once the transaction is complete. 

\item An \emph{delegation/revocation policy}~\cite{askarov2012, hicks05,
swamy06, matos2005} updates dynamically the sensitivity roles in a security
system to accommodate the mutable requirements of security, such as
delegating/revoking the access rights of a new/leaving employee. 

\end{itemize}

Moreover, there are a few case studies on the needed security properties in the
light of one specific context or task~\cite{mentalpoker, hicks2006,
preibusch2011, stoughton2014}, and build systems that provably enforces some
variants of declassification policy (e.g., CoCon~\cite{cocon},
CosMeDis~\cite{cosmedis}) and erasure policy (e.g., Civitas~\cite{civitas}).

Although the advances make it possible to specify and verify \emph{some
variants} of dynamic policy, cherry-picking the \emph{appropriate} policy is
still a daunting task: different policies (even when they belong to the same
kind) have very different syntax for specifying how a policy
changes~\cite{sabelfeld05}, very different nature of the security conditions
(i.e., noninterference, bisimulation and epistemic~\cite{broberg15}) and even
completely inconsistent notion of security (i.e., policies might disagree on
whether a program is secure or not~\cite{broberg15}). So even for veteran
researchers in information flow security, understanding the subtleties in the
syntax and semantics of each policy is difficult, evidenced by highly-cited
papers that synthesize existing knowledge on declassification
policy~\cite{sabelfeld05} and dynamic policy~\cite{broberg15}.  Arguably, it is
currently \emph{impossible} for a system developer/user to navigate in the
jungle of unconnected policies (even for the ones in the same category) when a
dynamic policy is needed~\cite{broberg15, sabelfeld05}.


\newsavebox\FlowExample
\begin{lrbox}{\FlowExample}
		\begin{minipage}{0.3\textwidth}
			\begin{lstlisting}[numbers=left,xleftmargin=0pt,framexleftmargin=0pt]
//reviews are declassified 
//after notiDate
$info := title + authors;$
$ date := curDate; $
$\If$ (date > notiDate) $\Then$
	 $ decision:= reviews  $
$\Else $ 
	 $ decision:= "" $
$\cod{output}$($ A $, info+decision);
\end{lstlisting}
		\end{minipage}
	\end{lrbox}
	
	\newsavebox\FlowExampleTrans
	\begin{lrbox}{\FlowExampleTrans}
		\begin{minipage}{0.3\textwidth}
\begin{lstlisting}[numbers=left,xleftmargin=0pt,framexleftmargin=15pt]
$ done := 0; $
$ mechant := $credit_card;  
$\cod{output}$($ M $, credit_card);
$ \cod{transaction}$();
$ done :=1 $; 
//mechant can not reuse 
//credit card any more
$ mechant : = 0 $ ; //erasure
//insecure: 
// $\cod{output}$($ M $, credit_card); 
\end{lstlisting}
			\end{minipage}
		\end{lrbox}
		
		
\newsavebox\DependExample
\begin{lrbox}{\DependExample}
	\begin{minipage}{0.25\textwidth}
\begin{lstlisting}[numbers=left,xleftmargin=0pt,framexleftmargin=15pt]
// John get access
//when join the company
state := employed;
John := resource1;
$\cod{output}$($ J $, resource1);
// Revoke John's access
state := left;
//insecure:
// $\cod{output}$($ J $, resource2); 
		\end{lstlisting}
	\end{minipage}
\end{lrbox}
		
\newcommand{\boxwidth}{0.27}
\newcommand{\leftmarginvalue}{25}
\newsavebox\DecSec
\begin{lrbox}{\DecSec}
\begin{minipage}{\boxwidth \textwidth}
\begin{lstlisting}[numbers=left,xleftmargin=\leftmarginvalue 
pt,framexleftmargin=15pt]
$ //~bid: \High $
$ submit:=bid $;
$ \outcmd{submit}{\High} $;
$ //~bid: \Low $
$ \outcmd{submit}{\Low} $;
\end{lstlisting}
\end{minipage}
\end{lrbox}

\newsavebox\DecInSec
\begin{lrbox}{\DecInSec}
\begin{minipage}{\boxwidth \textwidth}
\begin{lstlisting}[numbers=left,xleftmargin=\leftmarginvalue  
pt,framexleftmargin=15pt]
$ //~bid: \High $
$ submit:=bid $;
$ \outcmd{submit}{\Low} $;
$ //~bid: \Low $
$ \outcmd{submit}{\Low} $;
\end{lstlisting}
\end{minipage}
\end{lrbox}

\newsavebox\EraSec
\begin{lrbox}{\EraSec}
\begin{minipage}{\boxwidth \textwidth}
\begin{lstlisting}[numbers=left,xleftmargin=\leftmarginvalue  
pt,framexleftmargin=15pt]
$ //~credit\_card: \cod{M} $
$ copy:= credit\_card$;
$ \outcmd{copy}{\cod{M}} $;
$ //~credit\_card: \top $
$ copy:= 0 $;
$ \outcmd{copy}{\cod{M}} $;
\end{lstlisting}
\end{minipage}
\end{lrbox}

\newsavebox\EraInSec
\begin{lrbox}{\EraInSec}
\begin{minipage}{\boxwidth \textwidth}
\begin{lstlisting}[numbers=left,xleftmargin=\leftmarginvalue  
pt,framexleftmargin=15pt]
$ //~credit\_card: \cod{M} $
$ copy:= credit\_card$
$ \outcmd{copy}{\cod{M}} $;
$ //~credit\_card: \top $
$ //$ No Clear Up
$ \outcmd{copy}{\cod{M}} $;
\end{lstlisting}
\end{minipage}
\end{lrbox}

\newsavebox\RevSec
\begin{lrbox}{\RevSec}
\begin{minipage}{\boxwidth\textwidth}
\begin{lstlisting}[numbers=left,xleftmargin=\leftmarginvalue  
pt,framexleftmargin=\leftmarginvalue pt]
$ //book: \cod{bk}, notes: \cod{Alice}  $
$ //\cod{bk} \rightarrow \cod{Alice}  $
$ notes:= half(book)$;
$ \outcmd{notes}{\cod{Alice}} $;
$ //\cod{bk} \not\rightarrow \cod{Alice}  $
$ \outcmd{notes}{\cod{Alice}} $;
\end{lstlisting}
\end{minipage}
\end{lrbox}

\newsavebox\RevInSec
\begin{lrbox}{\RevInSec}
\begin{minipage}{\boxwidth\textwidth}
\begin{lstlisting}[numbers=left,xleftmargin=\leftmarginvalue  
pt,framexleftmargin=\leftmarginvalue pt]
$ //book: \cod{bk}, notes: \cod{Alice}  $
$ //\cod{bk} \rightarrow \cod{Alice}  $
$ notes:= half(book)$;
$ \outcmd{notes}{\cod{Alice}} $;
$ //\cod{bk} \not\rightarrow \cod{Alice}  $
$ \outcmd{book}{\cod{Alice}} $;
\end{lstlisting}
\end{minipage}
\end{lrbox}

\newsavebox\DepSec
\begin{lrbox}{\DepSec}
\begin{minipage}{0.22\textwidth}
\begin{lstlisting}[numbers=left,xleftmargin=15pt,framexleftmargin=15pt]
$ //~book: \{M\} $
$ //~card: \{Alice, M\} $;
$ \outcmd{card}{M} $;
$ //~book: \{M, Alice\}$ 
$ //~card: \{Alice\} $; 
$ \outcmd{book}{Alice} $;
\end{lstlisting}
\end{minipage}
\end{lrbox}

\newsavebox\DepInSec
\begin{lrbox}{\DepInSec}
\begin{minipage}{0.22\textwidth}

\begin{lstlisting}[numbers=left,xleftmargin=15pt,framexleftmargin=15pt]
$ //~book: \{M. Alice\} $
$ //~card: \{Alice\} $;
$ notes:= half(book)$;
$ \outcmd{notes}{Alice} $;
$ //~book: \{M\}$ 
$ \outcmd{book}{Alice} $;
\end{lstlisting}
\end{minipage}
\end{lrbox}


\begin{figure*}
	\centering
\small
	\begin{tabular}{|c|c|c|}
		\hline
		&&\\
		\cellcolor{sec}\usebox\DecSec 
		& \cellcolor{sec}\usebox\EraSec 
		& \cellcolor{sec}\usebox\RevSec  
			\\ \secprog  &\secprog  & \secprog  \\ && \\[-1ex]
		\cellcolor{insec}\usebox\DecInSec 
		& \cellcolor{insec}\usebox\EraInSec 
		& \cellcolor{insec}\usebox\RevInSec 
		\\ \insecprog & \insecprog & \insecprog \\ &&\\[-2ex]
		\cellcolor{header}\tabcenter{A. Declassfication} 
		& \cellcolor{header}\tabcenter{B. Erasure} 
		&\cellcolor{header}\tabcenter{C. Delegate/Revoke} 
		\\\hline
	\end{tabular}
	\caption{Examples of Dynamic Policies.}
\vspace{-3ex}
	\label{fig:application}
\end{figure*}

In this paper, we take a top-down approach and propose \policyname{}, the first
information flow policy that enables declassification, erasure, delegation and
revocation at the same time. One important insight that we developed during the
process is that erasure and revocation both strengthen an information flow
policy, despite their very different syntax in existing work.  However, an
erasure policy by definition disallows the same information leaked in the past
(i.e., before erasure) to be released in the future, while most revocation
policies allow so. This motivates the introduction of two kinds of policies,
which we call \emph{persistent} and \emph{transient} policies. 
The distinction can be interpreted as
a type of information flow which is permitted by some definitions but not by
others, called facets~\cite{broberg15}.

Moreover, \policyname{} is built on a novel formalization framework that is
shown to subsume existing security conditions that are formalized in different
ways (e.g., noninterference, bisimulation and epistemic~\cite{broberg15}).  More
importantly, for the first time, the formalization framework allows us to make
apple-to-apple comparison among existing policies, which are incompatible before
(i.e., one cannot trivially convert one to another). Besides the
distinction between \emph{persistent} and \emph{transient} policies mentioned
earlier, we also notice that it is more challenging to define a
\emph{transient} policy (e.g., erasure), as it requires a definition of the
\emph{precise} knowledge gained from observing one output event, rather than the more
standard \emph{cumulative} knowledge that we see in existing \emph{persistent}
policies.
 
Finally, we built a new \anntrace{} benchmark for testing and understanding
variants of dynamic policies in general. The benchmark consists of examples with
dynamic policies from existing papers, as well as new subtle examples that we
created in the process of understanding dynamic policies. We implemented our
policy and existing policies, and found that \policyname{} is the only one that
is both applicable and correct on all examples.

To summarize, this paper makes the following contributions:
\begin{enumerate}
\item We present a language abstraction with concise yet expressive security
specification (Section~\ref{sec:syntax}) that allows us to specify various
existing dynamic policies, including declassification, erasure, delegation and
revocation.

\item We present a new policy \policyname{} (Section~\ref{sec:semantics}). The
new definition resolves a few subtle pitfalls that we found in existing
definitions, and its security condition handles transient and persistent
policies in a uniform way.

\item We generalize the novel formalization framework behind \policyname{} and
show that it, for the first time, allows us to compare and contrast various
dynamic policies at the semantic level (Section~\ref{sec:framework}). The
comparison leads to new insights that were not obvious in the past, such as
whether an existing policy is transient or persistent.

\item We build a new benchmark for testing and understanding dynamic policies,
and implemented our policy and existing ones (Section~\ref{sec:evaluation}).
Evaluation on the benchmark suggests that \policyname{} is the only one that is
both applicable and correct on all examples.

\end{enumerate}

\section{Background and Overview}

\subsection{Security Levels}
As standard in information flow security, we assume the existence of a set of
security levels $\mathbb{L}$, describing the intended confidentiality of
information\footnote{Since integrity is the dual of confidentiality,
we will assume confidentiality hereafter.}. For generality, we \emph{do not}
assume that all levels form a Denning-style lattice. For instance, delegation
and revocation typically use principals/roles (such as $\cod{Alice,Bob}$) where the
\emph{acts-for} relation on principals can change at run time. For simplicity,
we use the notation $\ell\in \lattice$ if all levels form a lattice $\lattice$,
rather than $L\in \mathbb{L}$. Moreover, we use $\Low$ (public), $\High$
(secret) to represent levels in a standard two-point lattice where
$\Low\sqsubset \High$ but $\High\not\sqsubset \Low$.

\subsection{Terminology}
\label{sec:background}

Some terms in dynamic policy are overloaded and used inconsistently in the
literature. For instance, declassification is sometimes confused with dynamic
policy~\cite{broberg15}. To avoid confusion, we first define the basic
terminology that we use throughout the paper.

\begin{Definition}[Dynamic (Information Flow) Policy]
An information flow policy is dynamic if it allows the sensitivity of
information to change during one execution of a program.
\end{Definition}

As standard, we say that a change of sensitivity is \emph{downgrading} (resp.
\emph{upgrading}) if it makes information less sensitive (resp. more
sensitive).

Next, we use the examples in Figure~\ref{fig:application} to introduce the
major kinds of dynamic policies in the literature. For readability, we use
informal security specification in comments for most examples in the paper; a
formal specification language is given in Section~\ref{sec:syntax}.

\paragraph{Declassification}
Given a Denning-style lattice $\Lattice$,
declassification occurs when a piece of information has its sensitivity level
$\ell_1$ downgraded to a lower sensitivity level $\ell_2$ (i.e., 
$\ell_2\sqsubset
\ell_1$). Consider Figure~\ref{fig:application}-A which
models an online bidding system.  When bidders submit their bids to the system
during the bidding phase, each bid is classified that no other bidders are
allowed to learn the information. When the bidding ends, the bids are public to
all bidders. In the secure program (i), the bid is only revealed to a public
channel with level $ \Low $ (Line 5) when bidding ends.  However, the insecure
program (ii) leaks the bid during the bidding phase (Line 3).

\paragraph{Erasure}
Given a Denning-style lattice $\Lattice$, information erasure occurs when a
piece of information has its sensitivity level $\ell_1$ \revise{upgraded to a
more restrictive sensitivity level, or an incomparable level $\ell_2$ (i.e.,
$\ell_2 \not \sqsubseteq \ell_1$)}. 
Moreover, when information is erased to level $\top$, the sensitive information
must be removed from the system as if it was never inputted into the system.
Figure~\ref{fig:application}-B is from a payment
system. The user of the system gives her credit card information to the
merchandiser (at level $\cod{M}$) as payment for her purchase. When the
transaction is done, the merchandiser is not allowed to retain/use the credit
card information for any other purpose (i.e., its level changes to $\top$).
The secure program (i) only uses the credit card information during the
transaction (Line 3), and any related information is erased after the
transaction (Line 5). The insecure program (ii), however, fails to protect the
credit card information after the transaction (Line 6).

\paragraph{Delegation and Revocation}
Delegation and revocation are typically used together, in a
principal/role-based system~\cite{Arden:2015csf, rolebased, myers2000}. In this
model, information is associated with principals/roles, and a dynamic policy is
specified as changes (i.e., add or remove) to the ``acts-for''
relationship on principals/roles. Figure~\ref{fig:application}-C is from a book
renting system, where its customers are allowed to read books during the renting
period. In this example, $\cod{Alice}$ acts-for $\cod{bk}$ ($ \cod{bk}
\rightarrow \cod{Alice}  $) before line 3. Hence, she is allowed to take notes
from the book. When the renting is over, the book is no longer accessible to
$\cod{Alice}$ ($ \cod{bk} \not\rightarrow \cod{Alice}  $), but the notes remain
accessible to $\cod{Alice}$. The secure program (i) allows the customer to get
their notes (Line 6) learned during the renting period. The insecure program
(ii) fails to protect the book (Line 6) after the renting is over.

\subsection{Overview}
\label{sec:overview}
We use Figure~\ref{fig:application} to highlight two major obstacles of
understanding/applying various kinds of dynamic policies.

First, we note that a delegation/revocation policy (Example C) and an erasure
policy (Example B) use different formats to model sensitivity change. A
delegation/revocation policy  attaches \emph{fixed} security levels to data
throughout program execution; policy change is modeled as changing the acts-for
relation on roles. On the other hand, an erasure policy uses a fixed lattice
throughout program execution; policy change is modeled as \emph{mutable}
security levels on data. These two examples are similar from policy change
perspective, as they are both upgrading policies. But due to the different
specification formats, their relation becomes obscure.

Second, we note that Example B.ii and C.i are semantically very similar: both
examples first read data when the policy allows so, and then try to access the
data again when the policy on data forbids so. However, B.ii is considered
\emph{insecure} according to an erasure policy, while C.i is considered
\emph{secure} according to a revocation policy. 
Even when we only consider policies of the same kind (e.g.,
delegation/revocation), such inconsistency in the security notion also exists,
which is called \emph{facets of dynamic policies}~\cite{broberg15}.

Broberg et al.~\cite{broberg15} have identified a few facets, but
identifying other differences among existing policies is extremely difficult,
as they are formalized in different nature (e.g., noninterference, bisimulation
and epistemic). We can peek at the semantics-level differences based on a few
examples, but an apple-to-apple comparison is still impossible at this point.

In this paper, we take a top-down approach that rethinks dynamic policy from
scratch. Instead of developing four kinds of policies seen in prior work, we
observe that there are only \emph{two essential building blocks} of a dynamic
policy: upgrading and downgrading. With an expressive specification
language syntax (Section~\ref{sec:syntax}), we show that in terms of upgrading
and downgrading sensitivity, declassification (resp. erasure) is the same as
delegation (resp. revocation). In terms of the formal security condition of
dynamic policy, we adopt the epistemic model~\cite{askarov2007} and
develop a formalization framework that can be informally understood as the
following security statement: 
\begin{quotation}
	A program $c$ is secure iff for any event $t$ produced by $c$, the
	``knowledge'' \revise{gained about secret by } learning $t$ is bounded 
	by what's allowed by the policy at $t$.
\end{quotation}
We note that a key challenge of a proper security definition for the statement above
is to properly define the ``knowledge'' of learning a \emph{single} event $t$.
During the process of developing the formal definition, we discovered a new
facet of upgrading policies; the difference is that whether an upgrading policy
automatically allows information leakage (after upgrading) when it has happened
in the past. Consequently, we precisely define the ``knowledge'' of learning a single event
and make semantics-level choices (called \emph{transient} and \emph{persistent}
respectively) of the new facet explicit in \policyname{}
(Section~\ref{sec:semantics}).

To compare and contrast various dynamic policies (including \policyname{}), we
cast existing policies into the formalization framework behind \policyname{}
(Section~\ref{sec:framework}).  We find that the semantics of erasure and
revocation are drastically different: erasure policy is \emph{transient} by
definition, and most revocation policies are \emph{persistent}. The
semantics-level difference sheds light on why Example B.ii and C.i have
inconsistent security under erasure and revocation policies, even though they
are similar programs. 

\section{Dynamic Policy Specification}
\label{sec:syntax}

We first present the syntax of an imperative language with its
security specification. Based on that,  we show that the policy specification
is powerful enough to describe declassification, erasure, delegation and
revocation policies. Finally, we define a few notations to be used throughout
the paper.

\subsection{Language Syntax and Security Specification}
\label{sec:specification-syntax}
\begin{figure}
\small
\begin{align*}
	&\text{Variables ($\Vars$)}     &\quad &x, y, z \\
        &\text{Events ($\SEvents$)}       &\quad &s  \\
	&\text{Expressions ($\SEntities$)}   &\quad &\expr ::=\; x \mid n \mid e~\op~e \\
	&\text{Commands}      &\quad &\cmd  ::=\; \Skip \mid \cmd_1;\cmd_2 \mid 
	    \assign{x}{\expr}  \mid \while{\expr}{\cmd} \\
	&         & \quad &\mid \ifcmd{\expr}{\cmd_1}{\cmd_2} \mid \outcmd{b}{e} \\
        &         & \quad &\mid \cod{EventOn}(s) \mid \cod{EventOff}(s) \\
&\text{Level Sets ($\AllLevels$) }     &\quad & L   \\
&\text{Security Labels ($\Labels$)}    &\quad & \lab ::=\; L ~|~cnd? \lab_1 \circ 
\lab_2 ;~~~~ \\
&\text{Conditions}       &\quad &cnd::=\; s ~|~e~|~cnd\AND cnd~\\
&      & \quad      &|~cnd \OR cnd ~|~ \neg cnd \\
&\text{Mutation Directions}     &\quad  &\circ ::=\; \xrightarrow 
~|~\xleftarrow~|~\leftrightarrows \\
&\text{Policy Specification}     &\quad  &\revise{\spec: \Vars \mapsto 
\Labels[\diamond]}\\
&\text{Policy Type}       & \quad &\diamond ::=\;  Tran~|~Per
\end{align*}
\vspace{-2ex}
\caption{Language Syntax with Security Specification.}
\vspace{-3ex}
\label{fig:while-syntax}
\end{figure}


In this paper, we use a simple imperative language with expressive security
specification, as shown in Figure~\ref{fig:while-syntax}. The language provides
standard features such as variables, assignments, sequential composition,
branches and loops. Other features are introduced for security:
\begin{itemize}
\item We explicitly model information release by a release command $ \outcmd{b}{e} $;
it reveals the value of expression $ e $ to an information channel with
security label $b$.\footnote{In the literature, it is also common to model
information release as updates to a memory portion visible to an attacker. This
can be modeled explicitly as requiring an assignment $x:=e$ where $x$ has label
$b$ to emit a release command $ \outcmd{b}{v} $.}

\item We introduce distinguished security events $ \SEvents $. An event $s\in
\SEvents$ is similar to a Boolean; \revise{we distinguish $s$ and $x$ in
the language syntax to ensure that security events} can only be set and unset
using distinguished commands $ \cod{EventOn}(s) $ and $ \cod{EventOff}(s) $,
which set $ s $ to $ \True $ and $ \False $ respectively.
%
We assume that all security events are initialized with $ \False $.  
\end{itemize}


\subsubsection{Sensitivity Levels} 
\label{sec:label-syntax}
For generality, we assume a predefined set $ \AllLevels $ of all security
levels, and use level set $\labelset\subseteq \AllLevels$ to specify data
sensitivity.  Intuitively, a level set $\labelset$ consists of a set of levels
where the associated information can flow to. Hence, $L_1$ is less restrictive
as $L_2$, written as $L_1\sqsubset L_2$ iff $L_2\subset L_1$, and
$L_1\sqsubseteq L_2$ iff $L_2\subseteq L_1$.

Although the use of level set is somewhat non-standard, we note that it
provides better generality compared with existing specifications, such as a
level from a Denning-style lattice~\cite{denning-lattice} or a role in
a role-based model~\cite{Arden:2015csf, rolebased, myers2000}.
\begin{itemize}
\item Denning-style lattice: let $\Lattice$ be a security lattice. We can
define $ \AllLevels $ and the level set that represents $\ell\in
\Lattice$ as follows:
\begin{equation}
\label{eq:denning}
\AllLevels = \{\ell ~|~ \ell\in \Lattice \};~L_{\ell} \defn 
\{\ell' \in \Lattice
~|~\ell \LEQ \ell' \}
\end{equation}
Consider a two-point lattice $ \{\Low, \High\}$ with $\Low \sqsubset \High $.
It can be written as the follows in our syntax:
\begin{equation*}
\label{eq:denning-idea}
\AllLevels \defn  \{\Low, \High\}; ~~
L_{\High} \defn \{\High\};~~
L_{\Low} \defn \{\Low, \High\};
\end{equation*}

\item Role-based model: let $\mathbb{P}$ be a set of principals/roles and
$\cod{actsfor}$ be an acts-for relation on roles. We can define $
\AllLevels $ and the level set that represents $P\in \mathbb{P}$ as
follows:
\begin{equation}
\label{eq:principalbased}
\AllLevels = \mathcal{P}(\mathbb{P});~L_P \defn \{P'\in 
\mathbb{P}
~|~P'~\cod{actsfor}~P \}
\end{equation}
Consider a model with two roles $ \cod{Alice} $ and $ \cod{Bob} $ with $
\cod{Alice}~ \actsfor~\cod{Bob} $ but not the other way around.  It can be
written as the follows in our syntax:
\begin{align*}
\label{eq:dlm-idea}
 \AllLevels \defn \{\cod{Alice}, \cod{Bob}\};~
 L_{\cod{Alice}} \defn \{\cod{Alice}\}; \\
 L_{\cod{Bob}} \defn  \{\cod{Alice}, \cod{Bob}\};
\end{align*}
\end{itemize}

\subsubsection{Sensitivity Mutation}
The core of specifying a dynamic policy is to define how data sensitivity
changes at run time. This is specified by a security label $ b $.

A label can simply be a level set $ L $, which represents immutable sensitivity
throughout program execution.  In general, a label has the form of $ cnd?\lab_1
\circ \lab_2 $ where:
\begin{itemize}
\item  A trigger condition $cnd$ specifies \emph{when} the sensitivity changes.
There are two basic kinds of trigger conditions: a security event $ s $ and a
(Boolean) program expression $ e $. A more complicated condition can be
constructed with logical operations on $ s $ and $ e $. \revise{We assume that
a type system checks that whenever $cnd$ is an expression $e$,  $e$ is of
the Boolean type.}

\item The mutation direction $\circ$ specifies \emph{how} the information flow
restriction changes. There are two one-time mutation directions:
$cnd?\lab_1 \rightarrow \lab_2$ (resp. $cnd?\lab_1 \leftarrow \lab_2$) allows a
\emph{one-time} sensitivity change from $\lab_1$ to $\lab_2$ (resp. $\lab_2$ to
$\lab_1$) the first time that $ cnd $ evaluates to $\False$ (resp. $\True$).
On the other hand, a two-way mutation $ cnd?\lab_1 \leftrightarrows
\lab_2 $ allows arbitrary number of changes between $\lab_1$ and $\lab_2$ whenever the
value of $ cnd $ flips.

\end{itemize}
\revise{
\subsubsection{Policy Specification} 
The information flow policy on a program is specified as a function from
variables $ \Vars $ to security labels $ \Labels $ and a policy type
$\diamond$. The policy type can either be transient, or persistent (formalized
in Section~\ref{sec:semantics}). }

\subsection{Expressiveness}
\label{sec:expressiveness}
Despite the simplicity of our language syntax and security specification, we
first show that all kinds of dynamic policies in Figure~\ref{fig:application}
can be concisely expressed.
Then, we discuss how the specification covers the well-known \emph{what,
who, where} and \emph{when} dimensions~\cite{sabelfeld05,
Sabelfeld:2009journal} of dynamic policies.\footnote{The original definitions
focus on declassification policy, but the dimensions are applicable for dynamic
policies as well.}
Finally, we show that the specification language is powerful enough
to encode Flow Locks~\cite{Broberg:2006esop} and its successor
Paralocks~\cite{Broberg:2010popl}, \revise{a well-known} meta policy language
for building expressive information flow policies.

\subsubsection{Examples}
\label{sec:examples}
We first encode the examples in Figure~\ref{fig:application}.

\paragraph{Declassification and Erasure}
Both policies specify sensitivity
changes as mutating security level of information from some level $\ell_1$ to
$\ell_2$, where both $\ell_1$ and $\ell_2$ are drawn from a Denning-style
lattice $ \Lattice $.  Such a
change can be specified as $L_{\ell_1}\rightarrow L_{\ell_2}$, where
$L_{\ell_1}$ and $L_{\ell_2}$ are the level sets representing $\ell_1$ and
$\ell_2$, as defined in Equation~(\ref{eq:denning}).

For example, the informal policy on $credit\_card$ in Figure~\ref{fig:application}-B can
be precisely specified as $ erase?\{\}\leftarrow \{\cod{M}\}~[Tran]$ (we will discuss why
erasure is a transient policy in Section~\ref{sec:semantics}) with the security 
command $ \cod{EventOn}(erase)$ being inserted to Line 4 to trigger the 
mutation. 
	
\paragraph{Delegation and revocation}	
Both policies specify sensitivity
changes as modifying the acts-for relationship on principals, such as $ 
\cod{Alice}$ and
$\cod{Bob}$. 
Such a change can be specified as the old and new sets of roles who acts-for 
the
owner, say $P$, of information. That is, a change from from 
$\cod{actsfor}_1$
to $\cod{actsfor}_2$ can be specified as $L_1\rightarrow L_2$, where $L_i 
\defn
\{P'\in \mathbb{P} ~|~P'~\cod{actsfor}_i~P \} $.

For example, the policy on $book$ in Figure~\ref{fig:application}-C can be
specified as $revoke?\{\}\leftarrow \{\cod{Alice}\}~[Per]$ (we will discuss why
revocation is a persistent policy in Section~\ref{sec:semantics}) with a
security command $ \cod{EventOn}(revoke) $ being inserted to Line 5 to trigger
the mutation.
\footnote{\revise{We note that our encoding requires all changes to the
acts-for relation to be \emph{anticipated}, whereas a general
delegation/revocation policy might also offer the flexibility of changing the
acts-for relation dynamically.}}

\subsubsection{Dimensions of dynamic policy~\cite{sabelfeld05,
Sabelfeld:2009journal}}
\label{sec:dimensions}

\paragraph{What}
The \emph{what} dimension regulates what information's sensitivity is changed.
%
Since the policy specification is defined at variable level, our language does
not fully support partial release, which only releases a part of a secret
(e.g., the parity of a secret) to a public domain.
%
However, we note that the language still has \emph{some} support of partial
release. Consider the example in Figure~\ref{fig:application}-C.i.
The policy allows partial value $ half(book) $ to be accessible by Alice after
Line 5, while the whole value of $ book $ is not. As shown in
Section~\ref{sec:examples}, the partial release of $ half(book) $ in this
example can be precisely expressed in our language. We 
leave the full support of partial release as future work.

Moreover, we emphasize that the policy specification regulates the sensitivity
on the \emph{original value} of the variable. For example, consider
$\Gamma(h)=\High,\Gamma(x)=s?\High\rightarrow \Low$ for program:
\[  x:=h;~\cod{EventOn}(s);~\outcmd{\Low}{x};\]
The policy on $x$ states that
its original value, rather than its value right before output (i.e., the value
of $h$), is declassified to $\Low$. Hence, the program is insecure.  Therefore,
the specification language rules out laundering
attacks~\cite{sabelfeld2003, sabelfeld05}, which launders secrets
not intended for declassification.
  
\paragraph{Where}
The \emph{where} dimension regulates
\emph{level locality} (where information may flow to) and
\emph{code locality} (where physically in the code that information's
sensitivity changes).
It is obvious that a label $ cnd?\lab_1 \circ \lab_2 $ declare where
information may flow to after a policy change, and the security event $ s $ with
the security commands $ \cod{EventOn}(s) $ and $ \cod{EventOff}(s) $ 
specify the code locations where sensitivity changes.
 
\paragraph{When} 
The \emph{when} dimension is a temporal dimension, pertaining to when
information's sensitivity changes. This is specified by the trigger condition
$cnd$. For example, a policy $(\cod{paid}?\Low\leftarrow \High)$ allows
associated information (e.g., software key) to be released when payment has
been received. This is an instance of ``Relative'' specification defined
in~\cite{sabelfeld05}.

\paragraph{Who}
The \emph{who} dimension specifies a \emph{principal/role}, who controls the
change of sensitivity; one example is the Decentralized Label Model
(DLM)~\cite{ml-ifc-97}, which explicitly defines ownership in security labels.
While our specification language does not explicitly define ownership, we show
next that it is expressive enough to encode Flow
Locks~\cite{Broberg:2006esop} and Paralocks~\cite{Broberg:2010popl}, which in
turn are expressive enough to encode DLM~\cite{Broberg:2010popl}. Hence, the
specification language also covers the who dimension to some 
extent.


\subsubsection{Encoding Flow Locks~\cite{Broberg:2006esop}}  
\label{sec:paralock-label}
Both Flow Locks~\cite{Broberg:2006esop} and its successor
Paralocks~\cite{Broberg:2010popl} introduce \emph{locks}, denoted as $ \lock $,
to construct dynamic policies.  Let ${\alllocks}$ be a set of locks, and
$\mathbb{P}$ be a set of principals.  A ``flow lock'' policy is specified with
the following components:
\begin{itemize}
	\item Flow locks in the form of $\lockset \sat P$ where $ \lockset 
	\subseteq 
	\alllocks$ is the 
	lock set for principal $ P\in \mathbb{P}$.
	
	\item Distinguished commands $\cod{open}(\lock), \cod{close}(\lock)$ that 
	open and close the lock  $\lock \in \alllocks$.
\end{itemize}

To simplify notation, we use $ \spec(x, P) = \lockset $ to denote the fact that
$ \{\lockset \sat P\}$ is part of the ``flow locks'' of $ x $.
Paralocks security is formalized as \emph{an extension}
of Gradual Release~\cite{askarov2007}. In particular, paralock security is defined based on
\emph{sub-security condition} for each hypothetical attacker 
$A=(P_A,\lockset_A)$
where $P_A\in \mathbb{P}$ and $\lockset_A\subseteq \alllocks$:
\begin{itemize}
	\item A variable is considered ``public'' for attacker $A$ when $ \spec(x, 
	P_A)
	= \lockset_x \subseteq \lockset_A$; otherwise, it is considered ``secret'' for attacker $A$.
	
	\item A ``release event'', in gradual release sense, is defined as a period 
	of
	program execution when the set of \emph{opened} locks $\lockset_{open} 
	\subseteq
	\lockset_A$.
\end{itemize}

Hence, for each concrete $A=(P_A,\lockset_A)$, we can encode Paralocks security
as follows:
\begin{itemize}
	\item  We define a security event $ s_{\lock} $ for each lock $ \lock \in 
	\lockset $
	and the lock command $ \cod{open}(\lock) $ (resp. $ \cod{close}(\lock) $) 
	is 
	converted
	to $ \cod{EventOn}(s_\lock) $ (resp. $ \cod{EventOff}(s_\lock))$.
	
	\item Let $\Gamma(x)=\{P_A\}$ when $ \spec(x, P_A) = \lockset_x \subseteq
	\lockset_A$; otherwise, $\Gamma(x)=\{\} $ (i.e., secret for $P_A$).  
	
	\item Following the encoding of gradual release, we define
	$\Gamma'(x)=cnd?\Gamma(x):\{P_A\}$ where $cnd \defn \neg \bigwedge_{\lock 
	\not 
		\in \lockset_A} s_\lock$, i.e., all locks not in $\lockset_A$ must be 
		currently 
	closed, which implies an output event (not a release event):  
	$\lockset_{open} \subseteq \lockset_A$; otherwise, for a release event, $ x 
	$ 
	is public to $ P_A $.
\end{itemize}

\newsavebox\ParalockExample
\begin{lrbox}{\ParalockExample}
\begin{minipage}{.32\columnwidth}
\begin{lstlisting}[xleftmargin=0pt,framexleftmargin=0pt]
// x: {D,N}$\Rightarrow$a
// y: {N}$\Rightarrow$a
// z: {}$\Rightarrow$a 
$\cod{open}$(D);
y:=x;
$\cod{close}$(D);
$\cod{open}$(N);
z:=y;
\end{lstlisting}
\end{minipage}
\end{lrbox}
	
\newsavebox\ParalockTransExample
\begin{lrbox}{\ParalockTransExample}
	\begin{minipage}{.58\columnwidth}
\begin{lstlisting}[xleftmargin=0pt,framexleftmargin=0pt]
// x: $ s_N?\{a\}\leftrightarrows \{\}$
// y: $ s_N?\{a\}\leftrightarrows \{\}$
// z: $\{a\}$
$\cod{EventOn}(s_D)$;
y:=x; $\outcmd{ s_N?\{a\}\leftrightarrows\{\}}{x}$
$\cod{EventOff}(s_D)$;
$\cod{EventOn}(s_N)$;
z:=y;$\outcmd{\{a\}}{z}$
\end{lstlisting}
\end{minipage}
\end{lrbox}
		
\begin{figure}
	\centering
	\begin{tabular}{|c|c|}
		\hline
		 \cellcolor{sec}\usebox\ParalockExample
		& \cellcolor{sec}\usebox\ParalockTransExample \\
		\hline
	\end{tabular}
	\caption{An Example of Encoding Paralock for $ A=\configs{a, \{D\}} $.}
	\label{fig:paraloctrans}
\end{figure}

As a concrete example, we show the original Paralocks code and its transform
code in Figure~\ref{fig:paraloctrans} for $A=\configs{a,\{D\}}$. We note that
under the encoding, the first assignment $y:=x$ is under a release event since
only lock $D$ is open, which is a subset of $\lockset_A=\{D\}$; both the output
channel and the value can be read by $a$. On the other hand, the second
assignment $z:=y$ is \emph{not} under a release event, as an opened lock $N$ is
not possessed by attacker $A$. This is also reflected by the encoding: while
the output channel is observable to $a$ unconditionally, the value of $y$ has
policy $\{\}$ at that point, as $s_N=\True$.

Hence, we can encode Paralocks by explicitly checking the security of each
transformed program for each $A$, and accept the program iff all transformed
programs are secure.

\subsection{Interpretation of Security Specification}
\label{sec:state}

Intuitively, the security specification in Figure~\ref{fig:while-syntax}
specifies at each program execution point, what is the sensitivity of the
associated information.  We formalize this as an interpretation function of the
label, denoted as $ \psemantics{b}{\trace} $, which takes in a label $ b $ and
a trace $ \trace $, and returns a level set $ L $ as information flow
restrictions at the \emph{end} of $ \trace $.

\paragraph{Execution trace}

As standard, we model program state, called memory $m$, as a mapping from
program variables and security events to their values. The small-step semantics
of the source language is mostly standard (hence omitted), with exception of
the output and security event commands:
\begin{mathpar}
\inferrule*[right=S-Output]
{\configTwo{\Mem}{e}\evalto v}
{\configTwo{\Mem}{\outcmd{b}{e}} \xrightarrow[]{\configTwo{v}{b}}
\configTwo{\Mem}{\Skip} }
\and
\inferrule*[right=S-Set]
{}
{\configTwo{\Mem}{\cod{EventOn}(s)} \To 
\configTwo{\Mem\Mupdate{s}{\True}}{\Skip} }
\and
\inferrule*[right=S-Unset]
{}
{\configTwo{\Mem}{\cod{EventOff}(s)} \To 
\configTwo{\Mem\Mupdate{s}{\False}}{\Skip} }
\end{mathpar}
The semantics records all output events, in the form of $\configTwo{v}{b}$,
during program execution, as these are the only information release events
during program execution. Moreover, the distinguished security events $s$ are treated as
boolean variables, which can only be set/unset by the security event commands.

Based on the small-step semantics, executing a program $c$ under initial memory
$\Mem$ produces an execution trace $\trace$ with potentially empty output
events:
\[ \configs{c, m} 
\xrightarrow[]{b_1, v_1} \configs{c_1, m_1}  \cdots 
\xrightarrow[]{b_n, v_n}{}  \configs{c_n, m_n} .\]

We use $ \trace^{[i]} $ to denote the configuration (i.e., a
pair of program and memory) after the $i$-th evaluation step in the $\trace$,
and $ \len{\trace} $ to denote the number of evaluation steps in the trace.
For example, $ \trace^{[0]} $ is always the initial state of the execution, $
\trace^{[\len{\trace}]} $ is the ending state of a terminating trace $\trace$.
We use $ \trace^{[:i]} $ (resp. $ \trace^{[i:]} $) to denote a prefix (resp.
postfix) subtrace of $ \tau $ from the initial state up to (starting from) the
$ i $-th evaluation step. We use $ \trace^{[i:j]} $ to denote the subtrace of
$ \trace $ between $i$-th and $j$-th (inclusive) evaluation steps. Finally, we
write $\tau_1 \pref \tau_2$ when $\tau_1$ is a prefix of $\tau_2$.

\begin{figure}
\small

\begin{align*}
\psemantics{L}{\trace} &= L \\
\psemantics{cnd? b_1 \rightarrow b_2}{\trace} &=
\begin{cases}
\psemantics{b_1}{\trace}, & \cod{first}(cnd, \tau, \False) = -1
\\
\psemantics{b_2}{\trace^{[i:]}}, &i = \cod{first}(cnd, \tau, \False) \geq 0
\end{cases}\\
\psemantics{cnd? b_1 \leftarrow b_2}{\trace} &= 
\begin{cases}
\psemantics{b_2}{\trace}, & \cod{first}(cnd, \tau, \True) = -1 \\
\psemantics{b_1}{\trace^{[i:]}}, & i = \cod{first}(cnd, \tau, \True) \geq 0
\end{cases}\\
\psemantics{cnd? b_1 \leftrightarrows b_2}{\trace} &=
\begin{cases}
\psemantics{b_1}{\trace^{[i+1:]}}, 
& i = \cod{last(cnd, \tau, \False)}\neq \len{\trace} \\
\psemantics{b_2}{\trace^{[i+1:]}}, 
& i = \cod{last(cnd, \tau, \True)}\neq \len{\trace}\\
\end{cases}
\end{align*}
where $ \cod{first}(cnd, \tau, bl) $ returns the first index of $
\tau $ such that $ cnd $ evaluates to $ bl $, or $ -1 $ if such an index
does not exist;
   $ \cod{last}(cnd, \tau, bl) $ returns the last index of 
$ \tau $ such that $ cnd $ evaluates to $ bl $, or $ -1 $ if such an index does not exist. 
\caption{Interpretation of Security Labels}
\vspace{-3ex}
\label{fig:label-semantics}

\end{figure}

\paragraph{Interpretation of labels} \label{sec:dynamic-spec} 
We formalize the label semantics $ \psemantics{b}{\trace} $ in
Figure~\ref{fig:label-semantics}. $ \psemantics{b}{\trace} $  returns a level
set $L$ that precisely specifies where the information with policy $b$ can flow
to at the end of trace $\trace$.
For a (static) level set $L$, its interpretation is simply $L$ regardless of
$\trace$. 

For more complicated labels, the semantics also considers the temporal
aspect of label changes. For example, a one-time mutation label $cnd? b_1
\rightarrow b_2$ allows a one-time sensitivity change from $b_1$ to $b_2$ when the
first time that $ cnd $ evaluates to $\False$. Hence, let $i$ be the
\emph{first} index of $\trace$ such that $cnd$ evaluates to $\False$. Then,
$\psemantics{cnd? b_1 \rightarrow b_2}{\trace}$ reduces to
$\psemantics{b_1}{\trace}$ when no such $i$ exists (i.e., $ cnd $ always
evaluates to $\True$ in $\trace$), and it reduces to
$\psemantics{b_2}{\trace^{[i:]}}$ otherwise.  Note that in the latter case, it
reduces to $\psemantics{b_2}{\trace^{[i:]}}$ rather than
$\psemantics{b_2}{\trace}$ to properly handle nested conditions: any nested
condition in $b_2$ can only be evaluated \emph{after} $cnd$ becomes $\False$.
The dual with $\leftarrow$ is defined in a similar way. Note that $cnd? b_1
\rightarrow b_2$ and $\neg cnd?b_2 \leftarrow b_1$ are semantically the same;
we introduce both for convenience.

Finally, the bi-directional label (with $\leftrightarrows$) is interpreted
purely based on the last configuration of $\trace$: \revise{let $i$ be the
\emph{last} index in $\trace$ such that $cnd$ evaluates to $\False$. Then,
$i\neq \len{\trace}$ implies that $ cnd $ evaluates to $ \True $ at the end of
$\tau$; hence, the label reduces to $b_1$. Note that $b_1$ is evaluated under
$\trace^{[i+1:]}$ in this case to properly handle (potentially) nested
conditions in $b_1$: any nested condition in $b_1$ can only be evaluated
\emph{after} $cnd$ becomes $\True$.
}

Moreover, we can derive a \emph{dynamic}
specification for each execution point $i$, written as $\gamma_i$, such that
\[\forall x.~ \gamma_i(x) = 
\psemantics{\spec(x)}{\trace_{[:i]}}\]

Additionally, we overload $ \gamma_i $ to track the dynamic interpretation of a 
label
$b$ for each execution point $i$:
\[
 \forall b.~ \gamma_i(b) = 
\psemantics{b}{\trace_{[:i]}} \]
To simplify notation, we write 
\[
\configs{c_0, m_0} \termout  \vec{t} 
\]
if the execution $\configs{c_0, m_0}$ \emph{terminates}\footnote{ \revise{In
this paper, we only consider output sequences $ \vec{t} $ produced by  $
\configs{c_0, m_0} \termout \vec{t} $. Hence, only the terminating executions
are considered in this paper, making our knowledge and security definitions in
Section~\ref{sec:semantics} termination-insensitive.} 
Termination sensitivity is an orthogonal issue
to the scope of this paper: dynamic policy.}
with an \emph{extended}
output sequence $ \vec{t} $, which consists of extended output events $ t \defn
\configs{b, v, \gamma}$, where $ b, v $ are the output events on $\tau$, and
$\gamma$ is the dynamic specification at the corresponding execution point.  We
use $ t.b $, $ t.v $ and $ t.\gamma $ to refer to each component in the
extended output event. We use the same index notation as in trace, where $ 
\vec{t}^{[i]} $ returns the $ i $-th output event, and $ \vec{t}^{[:i]} $ 
returns the prefix output sequence up to (included) the $ i $-th output. $ 
\vec{t}^{[:0]}$ returns an empty sequence.

\section{\policyname}
\label{sec:semantics}

In this section, we define \policyname{}, an end-to-end information flow policy
that allows information flow restrictions to downgrade and upgrade in arbitrary
ways.

%

\subsection{Semantics Notations}

\paragraph{Memory Closure}
For various reasons, we need to define a set of initial memories that are
indistinguishable from some memory $m$. Given a set of 
variables $X$, we
define the memory closure of $m$ to be a set of memory who agrees on the value
of each variable $x \in X$:
\begin{Definition}[Memory Closure] Given a memory $m$ and a set of
variables $X$, the memory closure of $m$ on $X$ is:
	\label{def:mem-eq}
	\begin{align*}
	\closure{m}_{X} \defn \{m' \mid \forall x\in X.~m(x)=m'(x)\}
	\end{align*}
\end{Definition}
For simplicity, we use the following short-hands: 
\begin{align*}
\closure{m}_{L, \gamma} &\defn  \closure{m}_{\{x~\mid   \gamma(x) \subseteq L
	\}} \\
\closure{m}_{\neq b} &\defn  \closure{m}_{\{x~\mid  \Gamma(x)\not =b 
	\}} 
\end{align*}
where $ \closure{m}_{L, \gamma} $ is the memory closure on all variables whose
sensitivity level is less or equally restrictive than a level $ L $ according to
$ \gamma $, and $ \closure{m}_{\neq b} $ is the memory closure on
variables whose security policy is not $b$: a set of memories whose
value only differ on variables with policy $b$.

\paragraph{Trace filter}
For various reasons, we need a filter on output traces to focus on relevant
subtraces (e.g., to filter out outputs that are not visible to an attacker).
Each trace filter can be defined as a Boolean function on $\configs{b,
	v,\gamma}$. With a filter function $f$ (that returns $\False$ for irrelevant
outputs), we define the projection of outputs as follows:
\begin{Definition}[Projection of Trace] 
	\label{def:proj}
	\begin{equation*}
	\proj{\vec{t}}_f \defn \configs{ \configs{b, v, \gamma} \in 
		\vec{t}~\mid~f(b, v, \gamma)}
	\end{equation*}
	
	We define the following short-hand for commonly used 
	filter, $ L $-projection filter, where the resulting trace consists of outputs 
	currently observable at level $ L $ :
	\begin{align*}
	\proj{\vec{t}}_L &\defn \proj{\vec{t}}_{\lambda b, v, \gamma.~\gamma(b) \subseteq 
		L} 
	\end{align*}
\end{Definition}

\subsection{Key Factors of Formalizing a Dynamic Policy}
\label{sec:challenges}
Before formalizing \policyname{}, we first introduce knowledge-based security
(i.e., epistemic security)~\cite{askarov2007}, which is widely used in
the context of dynamic policy. Our formalization is built on the following
informal security statement, \revise{which is motivated by~\cite{askarov2012}}:
\begin{quotation}
	A program $c$ is secure iff for any event $t$ produced by $c$, the
	``knowledge'' \revise{gained about secret by } observing $t$ is bounded by what's allowed by 
	the policy at
	$t$.
\end{quotation}

We first introduce a few building blocks to formalize ``knowledge'' and
``allowance'' (i.e., the allowed leakage).

\subsubsection{Indistinguishability}

A key component of information flow security is to define trace
indistinguishability: whether two program execution traces are distinguishable to an
attacker or not. Given an attacker at level set $L$, each release event
$\configs{b,v,\gamma}$ is visible iff $\gamma(b)\LEQ L$ by the attack model.
Hence, as standard, we define an indistinguishability relation, written as
$\sim_L$, on traces as
\[\sim_L\ \defn \{(\vec{t}_1,\vec{t}_2)~|~\proj{\vec{t}_1}_L \pref
\proj{\vec{t}_2}_L\}\]
Note that an attacker cannot
rule out any execution whose prefix matches $t_1$. Hence, the prefix relation
is used instead of identity.

\subsubsection{Knowledge gained from observation}

Following the original definition of knowledge in~\cite{askarov2007},
we define the knowledge gained by an attacker at level set $L$ via observing a
trace $\vec{t}$ produced by a program $c$ as:\footnote{We slightly modified the original
definition to exclude ``initial knowledge'', the attacker's knowledge before
executing the program.}
\begin{equation}
\label{eq:knowledge}
k_1(c, \vec{t}, L) \defn \{ m~\mid~
\configs{c, m} \termout \vec{t'} \AND \vec{t}\sim_L
\vec{t'}
\} 
\end{equation}
Intuitively, it states that if one initial memory $m$ produces a trace that is
indistinguishable from $\vec{t}$, then the attacker cannot rule out $m$ as one
possible initial memory. Note that by definition, the smaller the knowledge set
is, the more information (knowledge) is revealed to the attacker.

\revise{Recall that by definition, $\configs{c, m} \termout \vec{t'}$ only
considers \emph{terminating} program executions. Hence, the knowledge
definition above is the \emph{termination-insensitive} version of knowledge
defined in~\cite{askarov2007}. As a consequence, the security semantics that we
define in this paper is also termination-insensitive.}

\subsubsection{Policy Allowance} 
To formalize security, we also need to define for each output event $t$ on a
trace, what is the allowed leakage \revise{to an attacker at a level set $L$}.
As knowledge, policy allowance, written as $\allowence(m,\vec{t},b,L)$, is
defined as a set of memories that should remain indistinguishable to the actual
initial memory $m$ at the end of output sequence $\vec{t}$.

Consider a dynamic label $b\in \Labels$, memory $m$ and output sequence $\vec{t}$ of
interest, as well as an attacker at level $L$, we can define policy allowance
as follows:
\[
\allowence(m,\vec{t},b,L)\defn \closure{m}_{\neq b}
\]
\revise{
Intuitively, it specifies the initial knowledge of an attacker at level set
$L$: the attacker cannot distinguish any value difference among variables with
the dynamic label $b$. Thus, any variable with the label $b$ is initially
indistinguishable to the attacker.  Eventually, \policyname{} checks that for
each label $b\in \Labels$, gained knowledge is bounded by the allowance with
respect to $b$. Hence, the security of each variable is checked.}

\subsection{Challenges of Formalizing a General Dynamic Policy}
We next show that it is a challenging task to formalize the security of a
general-purpose dynamic policy that allows downgrading and upgrading to occur
in arbitrary ways.

\subsubsection*{Challenge 1: \revise{Permitting both increasing and decreasing
knowledge}} 
\revise{Allowing both downgrading and upgrading in arbitrary ways means that
our general policy must permit reasoning about both increasing knowledge (as in
declassification) and decreasing knowledge (as in erasure).} While
Equation~\ref{eq:knowledge} and its variants are widely used to formalize
declassification policy~\cite{askarov2007,Broberg:2010popl}, they \emph{cannot}
reason about increasing knowledge. 
\revise{For example, it is easy to check that for any $c, \vec{t}, \vec{t}',
L$, we have 
\[\vec{t}\pref \vec{t'} \Rightarrow k_1(c, \vec{t}, L)\supseteq k_1(c, \vec{t'}, L)\]
according to Equation 3.
}
As other variants, the knowledge set $k_1$ is monotonically decreasing (hence,
the knowledge that it represents is increasing by definition) as more events on
the same execution are revealed to an attacker~\cite{askarov2007, askarov2012,
Delft:2015post}. 

\revise{However, we need to reason about decreasing knowledge for an erasure
policy.} Consider the example in Figure~\ref{fig:application}-B, where the
value of credit card is revealed by the first output at Line 3. Given any
program execution $\configs{c, m} \termout \vec{t}$, we have $k_1(c,
\vec{t}^{[:i]}, M)=\{m\}$ for all $i\geq 1$. However, as the sensitivity of $
credit\_card $ upgrades from $ M $ to $ \top $ when $i=2$ (i.e., the second
output), the secure program (i) can be incorrectly rejected: $k_1(c,
\vec{t}^{[:2]}, M)=\{m\}$ means that the value of $ credit\_card $ is known to
the attacker, which violates the erasure policy at that point.

\textbf{Observation 1.} Equation~\ref{eq:knowledge} is not suitable for an
upgrading policy, since \revise{it fails to reason about decreasing knowledge.}
The issue is that knowledge gained from $\vec{t}$ is defined as 
the full knowledge gained from observing \emph{all} outputs on $\vec{t}$. \revise{Return
to the secure program in Figure~\ref{fig:application}-B.i. We note that the first
and second outputs together reveal the value of $ credit\_card $, but the
second event alone reveals no information, as it always outputs 0. Hence, we
can precisely define the \emph{exact knowledge} gained from learning \emph{each}
output to permit both increasing and decreasing knowledge.}

\subsubsection*{Challenge 2: Indistinguishability $\sim_L$ is
\revise{inadequate for a general dynamic policy}}
\revise{
As shown earlier, indistinguishability $\sim_L$ is an important component of a
knowledge definition; intuitively, by observing an execution $\configs{c, m}
\termout \vec{t}$, an attacker at level set $L$ can rule out any initial memory
$m'$ where $m \not\sim_L m'$ (i.e., $m'\not\in k_1(c,\vec{t},L)$). However, the
naive definition of $\sim_L$ might be inadequate for declassified outputs.}
Consider the following secure program, where $x$ is first
downgraded to $\Low$ and then upgraded to $\High$.

\newsavebox\FilteringFExample
\begin{lrbox}{\FilteringFExample}
	\begin{minipage}{0.25\textwidth}
		\begin{lstlisting}[xleftmargin=15pt,framexleftmargin=15pt, 
		numbers=left]
// x : $\Low$
$ \If $ (x>0) $\outcmd{\Low}{1}$;
$\outcmd{\Low}{1}$
// x : $\High$
$\outcmd{\Low}{2}$
\end{lstlisting}
\end{minipage}
\end{lrbox}
\begin{tabular}{|c|}
	\hline
	\cellcolor{sec}\usebox\FilteringFExample\\ 
	\hline
\end{tabular}

Note that the program is secure since the only output when $x$ is secret
reveals a constant value. \revise{Assume that the initial value of $x$ is
either 0 or 1. Hence, there are two possible executions of the program with $ 
\gamma_1(x)=\Low $ and $ \gamma_2(x)=\High $:}
{\small
\begin{align*}
\configs{c, m_1} &\termout \configs{\Low,1,\gamma_1}\cdot 
\configs{\Low,2,\gamma_2} \\
\configs{c, m_2} &\termout \configs{\Low,1,\gamma_1}\cdot 
\configs{\Low,1,\gamma_1}\cdot \configs{\Low,2,\gamma_2}
\end{align*}
}
\revise{The issue is in the first execution. By observing the first output, an
attacker at $\Low$ cannot tell if the execution starts from $m_1$ or $m_2$, as
both of them first output 1. However, the attacker can rule out $m_2$ by
observing the second output with the value of 2. Note that the change of knowledge
(from $\{m_1,m_2\}$ to $\{m_1\}$) violates the dynamic policy governing the
second output: the policy on $x$ is $\High$, which prohibits the learning of
the initial value of $x$.}

\textbf{Observation 2.}
\revise{The inadequacy of relation $\sim_L$ roots from the fact that, due to
downgrading, the public outputs of different executions might have various
lengths. Therefore, outputs at the same index but produced by different
executions might be incomparable. To resolve the issue,}
we observe that any information release (of $x$) when $x$ is $\Low$ is
\emph{ineffective}, in the sense that the restriction on $x$ is not in effect.
In the example above, the outputs with value $1$ are all ineffective, as $ x $
is public when the outputs at lines 2 and 3 are produced.
This observation motivates the secret projection filter, which finds out the
\emph{effective} outputs for a given secret. 
\begin{Definition}[Secret Projection of Trace] 
Given a policy $b$ and an attacker at level $L$, a secret projection of trace
is a subtrace where information with policy $b$ cannot flow to $L$ and the
output channel is visible to $ L $ :
	\begin{align*}
	\proj{\vec{t}}_{b, L} &\defn \proj{\vec{t}}_{\lambda b', n, \gamma.~\gamma(b)  
	\not \subseteq L~\AND ~\gamma(b') \subseteq L } 
	\end{align*}
\end{Definition}

Return to the example above, the effective subtraces starting from $m_1$ and
$m_2$ are both $\configs{\Low,2,\gamma_2(x)=\High}$, which remains
indistinguishable to an attacker at level $\Low$.

\subsubsection*{Challenge 3: Effectiveness is also inadequate}
With Observation 2, it might be attempting to define indistinguishability based
on $\proj{\vec{t}}_{b, L}$, rather than $\proj{\vec{t}}_{L}$. However, doing so
is problematic as shown by the following program.

\newsavebox\FilteringBExample
\begin{lrbox}{\FilteringBExample}
	\begin{minipage}{0.3\textwidth}
		\begin{lstlisting}[xleftmargin=15pt,framexleftmargin=0pt, 
		numbers=left]
  // x : $\High$
  $ \If $ (x>0) $\cod{output}$($ \Low $, 1);
  // x : $\Low$
  $ \If $ (x<=0) $\cod{output}$($ \Low $, 1);
		\end{lstlisting}
	\end{minipage}
\end{lrbox}
\begin{tabular}{|c|}
	\hline
	\cellcolor{sec}\usebox\FilteringBExample\\ 
	\hline
\end{tabular}


With two initial memories $m_1(x)=0, m_2(x)=1$, we have
\begin{align*}
\configs{c, m_1} &\termout \configs{\Low,1,\gamma_1(x)=\Low} \\
\configs{c, m_2} &\termout \configs{\Low,1,\gamma_2(x)=\High}
\end{align*}
Note that only the value of $x$ is revealed on the public channel. Hence, the
program is secure as it always outputs $1$. However, the effective subtrace
starting from $m_1$ is $\emptyset$ and that starting from $m_2$ is
$\configs{\Low,1,\gamma_2(x)=\High}$, suggesting that the program is insecure:
the value of $x$ is revealed by the first output from $m_2$, while the policy
at that point ($\High$) disallows so.

\textbf{Observation 3.} We note that both indistinguiability and effectiveness
are important building blocks of a general-purpose dynamic policy. However, the
challenge is how to combine them in a meaningful way. \revise{We will build our
security definition on both concepts and justify why the new definition is
meaningful in Section~\ref{sec:dynrelease}.}

\subsubsection*{Challenge 4: Transient vs. Persistent Policy} So far, the
policy allowance $\allowence(m)$ ignores what information has been leaked in
the past. However, in the persistent case such as
Figure~\ref{fig:application}-C, the learned information ($ note $) remains
accessible even after the policy on $ book $ upgrades. In general, we define
transient and persistent policy as:

\begin{Definition}[Transient and Persistent Policy]
A dynamic security policy is persistent if it always allows to reveal
information that has been revealed in the past. Otherwise, the policy is
transient.
\end{Definition}

\textbf{Observation 4.} Both transient and persistent policy have real-world
application scenarios. Hence, a general-purpose dynamic policy should support
both kinds of policies, in a unified way.

\subsection{\policyname}
\label{sec:dynrelease}
We have introduced all ingredients to formalize \policyname, a novel
end-to-end, general-purpose dynamic policy. 

\revise{To tackle the challenges above, we first formalize the attacker's knowledge
gained by observing the \emph{last} event $t'$ on a trace $\vec{t}\cdot t'$.
Note that simply computing the knowledge difference between observing
$\vec{t}\cdot t'$ and observing $\vec{t}$ does not work. Consider
the example in Figure~\ref{fig:application}-B.ii. Given any program execution
$\configs{c, m} \termout \vec{t}$, we have $k_1(c, \vec{t}^{[:i]}, M)=\{m\}$
for all $i\geq 1$. Hence, the difference between the knowledge gained with or
without the output at Line 6 is $\emptyset$, suggesting that no knowledge is
gained by observing the output at Line 6 alone, which is incorrect as it
reveals the credit card number.}

Instead, we take inspiration from probabilities to formalize the attacker's
knowledge gained by observing a single event on a trace. \revise{Consider a
program $c$ that produces the following sequences of numbers give the
corresponding inputs: 
\begin{align*}
\text{ input 1: } s_1 &=(1\cdot 1 \cdot 3) \\
\text{ input 2: } s_2 &=(2\cdot 2 \cdot 3) \\
\text{ input 3: } s_3 &=(1\cdot 1 \cdot 3) \\
\text{ input 4: } s_4 &=(2\cdot 2 \cdot 2) 
\end{align*}
Consider the following question: what is the probability that the program
generates a sequence where the last number is identical to the last number of
$s_1$? Obviously, besides $s_1$, we also need to consider sequences $s_2$ and
$s_3$ since albeit a different sequence, $s_2$ is \emph{consistent} with $s_1$
in the sense that the last output is $3$, and $s_3$ is \emph{indistinguishable}
(i.e., identical) to $s_1$. More precisely, we can compute the probability as
follows:
\[\Sigma_{s \in \cod{consist}(s_1)} P(s)\]
where the consistent set $\cod{consist}(s_1)$ is the set of
sequences that produce the same last number as $s_1$, i.e., $\{(1\cdot 1 \cdot
3), (2\cdot 2 \cdot 3)\}$. Assuming a uniform distribution on program inputs,
we have that the probability is $P(1\cdot 1 \cdot 3)+P(2\cdot 2 \cdot
3)=(0.25+0.25)+0.25=0.75$. Note that the indistinguishable sequences $s_1$ and
$s_3$ are implicitly accounted for in $P(1\cdot 1 \cdot 3)$.}

To compute the knowledge associated with the \emph{last} event on a trace
$\vec{t}$, we first use effectiveness to identify \emph{consistent} traces
whose last event on the effective subset is the same:
\begin{Definition}[Consistency Relation] 
	\label{def:consistency}
Two output sequences $\vec{t}_1$ and $\vec{t}_2$ are consistent w.r.t.
a policy $b$ and an attack level $L$, written as $\vec{t_1} \equiv_{b,L} \vec{t_2}$ if
	\begin{align*}
	n=\len{\proj{\vec{t}_1}_{b,L}}=\len{\proj{\vec{t}_2}_{b,L}} \land
	\proj{\vec{t}_1}_{b,L}^{[n]}=\proj{\vec{t}_2}_{b,L}^{[n]}
	\end{align*}
\end{Definition}

\revise{Note that despite the extra complicity due to trace projection, the
consistency relation is similar to the consistent set $\cod{consist}(s_1)$ in
the probability computation example. Next, we define the precise knowledge
gained from the last event of $\vec{t}$ based on both the consistency relation
and knowledge. Note that since knowledge is a set of
memories, rather than a number, the summation in the probability case is
replaced by a set union. Similar to the probability of observing each sequence,
the knowledge $k_1$ also implicitly accounts for all indistinguishable traces
(Equation~\ref{eq:knowledge}).}

\begin{Definition}[Attacker's Knowledge Gained from the Last Event] For an attacker at level
set $ L $, the attacker's knowledge w.r.t. information with policy $b$, after
observing the last event of an output sequence $ \vec{t} $ of program $ c $, is
the set of all initial memories that produce an output sequence that is
indistinguishable to some consistent counterpart of $\vec{t}$: 
\label{def:eff-knowledge}
\begin{align*}
k_2(c, \vec{t}, L, b) = \bigcup_{\exists m',j.~\configs{c, m'} \termout 
\vec{t'} \AND t'^{[:j]} \equiv_{b,L}  \vec{t} } k_1(c, 
\vec{t'}^{[:j]}, 
L )
\end{align*}
\end{Definition}

\revise{
To see how Definition~\ref{def:eff-knowledge} tackles Challenges 2 and 3, we
revisit the code example under each challenge.
\begin{itemize}
\item Challenge 2: Recall that with $m_1(x)=0$, $m_2(x)=1$, $ \gamma_1(x)=\Low 
$ and $ \gamma_2(x)=\High $, there are two execution traces
{\small
\begin{align*}
\configs{c, m_1} &\termout \configs{\Low,1,\gamma_1}\cdot 
\configs{\Low,2,\gamma_2} \\
\configs{c, m_2} &\termout \configs{\Low,1,\gamma_1}\cdot 
\configs{\Low,1,\gamma_1}\cdot \configs{\Low,2,\gamma_2}
\end{align*}}
It is easy to check that the two output
sequences are consistent according to Definition~\ref{def:consistency}. Hence,
in both traces, the knowledge gained from the last output is $\{m_0,m_1\}$, due
to the big union in $k_2$. Hence, we correctly conclude that no information is
leaked by the last output in both traces.
\item Challenge 3: Recall that with $m_1(x)=0$, $m_2(x)=1$, $ \gamma_1(x)=\Low 
$ and $ \gamma_2(x)=\High $, there are two
execution traces
{\small
\begin{align*}
\configs{c, m_1} &\termout \configs{\Low,1,\gamma_1} \\
\configs{c, m_2} &\termout \configs{\Low,1,\gamma_2}
\end{align*}}
While the two traces are not consistent with each other, we know that $k_1(c,
\configs{\Low,1,\gamma_2(x)=\High}, \Low)=\{\configs{\Low,1,\gamma_1(x)=\Low},
\configs{\Low,1,\gamma_2(x)=\High}\}$ since the two traces satisfy
$\sim_{\Low}$. Hence, the knowledge gained from the last event is
$\{m_0,m_1\}$, and we correctly conclude that no information is leaked by the
last output.
\end{itemize}
}

\if
To see how Definition~\ref{def:eff-knowledge} 
captures the \emph{effective 
leakage}. Let's use the example below. It is a secure program, 
combining Figure~\ref{fig:indistinguish} (Line 1-4) and the erasure program in 
Figure~\ref{fig:application}-B(i) (Line 5-7).

\newsavebox\FilteringCExample
\begin{lrbox}{\FilteringCExample}
	\begin{minipage}{0.3\textwidth}
		\begin{lstlisting}[xleftmargin=15pt,framexleftmargin=0pt, 
		numbers=left]
// x : H
$ \If $ (x>0) $\cod{output}$($ L $, 1);
// x : L
$ \If $ (x<=0) $\cod{output}$($ L $, 1);
$\cod{output}$($ L $, x);
// x : H
$\cod{output}$($ L $, 0);
		\end{lstlisting}
	\end{minipage}
\end{lrbox}
\hspace*{0.2cm}
\begin{tabular}{|c|}
	\hline
	\cellcolor{sec}\usebox\FilteringCExample\\ 
	\hline
\end{tabular}

All the executions of this program generates three outputs $ 1\cdot x \cdot 0 $ 
at level $ L $. There are four outputs specified in the program and two of 
them, at Line 2 and Line 7,  are effective for the secret of $ x $. 
\begin{itemize}
	\item $ k_2(c, \vec{t}^{[:1]}, L, x)  = \mathbb{U} $. 
	Attacker's effective
	knowledge of secret $ x $ after observing the first output is full set:~~if 
	$ x>0 $, $ m' $ can 
	be any memory with $ x>0 $ and $ j=1 $, then $ k_1 $ from 
	Equation~\ref{eq:knowledge} returns full set; otherwise, if $ x\le 0 $, $ 
	\proj{\vec{t}}_{x, L} $ returns empty trace, then $ k_1 $ with empty trace 
	is trivially full set.
	Oppositely, the effect leakage at this point is empty. The secret of the $ 
	x $ is not leaked. 
	\item $ k_2(c, \vec{t}^{[:2]}, L, x)  = \mathbb{U} = k_2(c, \vec{t}^{[:3]}, 
	L, x) $. 
	The effective attacker's knowledge on $ x $ after observing two outputs, 
	or all three outputs,  are still full set, 
	which goes through the same computation as the first case since the second 
	output is not effective, and the third output is a 
	constant, not related to $ x $. 
\end{itemize}
This example can not be correctly verified if using only the indistinguishable 
attacker's knowledge as in Equation~\ref{eq:knowledge}, or only the effective 
output as in Definition~\ref{def:effective}. 
\fi

To tackle Challenge 4, we observe that a persistent policy allows information
leaked in the past to be released again, while a transient policy disallows so.
This is made precise by the following refinement of policy allowance:
\begin{equation}
\label{eq:allowance}
\allowence(m,\vec{t},b,L) \defn
\begin{cases} 
\closure{m}_{\neq b}, & b \text{ is transient} \\
\closure{m}_{\neq b} \cap k_1(c, \vec{t}^{[:{\len{\vec{t}}-1}]},L), & b \text{ 
	is persistent}
\end{cases}
\end{equation}
where $k_1(c, \vec{t}^{[:{\len{\vec{t}}-1}]},L)$ is the knowledge from every output event
in $\vec{t}$ except the last one. Note that since the knowledge here represents
the cumulative knowledge gained from observing all events, we use the standard
knowledge $k_1$ instead of the knowledge gained from the last event $k_2$ here.

Putting everything together, we have \policyname{} security, where for any
output of the  program, the attacker's knowledge gained from observing the
output is always bounded by the policy allowance at that output point.
\begin{Definition}[\policyname] 
\label{def:newpolicy}
	\begin{multline*}
	\forall m, L\subseteq \mathbb{L}, b \in  \mathbb{B}, 
	\vec{t}.~\configs{c,
		m} \termout \vec{t}  
	\implies 
	\forall 1\leq i\leq \len{\vec{t}}.\quad \\
	k_2(c,\vec{t}^{[:i]}, L, b) 
	\supseteq 
	\begin{cases} 
	\closure{m}_{\neq b}, & \text{transient} \\
	\closure{m}_{\neq b} \cap k_1(c, \vec{t}^{[:{i-1}]},L), & 
	\text{persistent}
	\end{cases}
 	\end{multline*}
\end{Definition}

\section{Semantics Framework For Dynamic Policy}
\label{sec:framework}

While various forms of formal policy semantics exist in the literature,
different policies have very different nature of the security conditions (i.e.,
noninterference, bisimulation and epistemic~\cite{broberg15}).  In this section,
we generalize the formalization of \policyname{}
(Definition~\ref{def:newpolicy}) by abstracting away its key building blocks.
Then we convert various existing dynamic policies into the
formalization framework and provide the first apple-to-apple comparison
between those policies. 

\begin{table*}
\centering
\scalebox{0.85}{
\begin{tabular}{|l|l|l|l|}
\hline
&  $\sim(\vec{t}_1,\vec{t}_2)$
& $\allowence(m,\vec{t},b,L)$, $ i = \len{\vec{t}} $
& $\equiv(\vec{t}_1,\vec{t}_2)$ 
 \\ \hline
\ifstaticni
\leftcentering{Static Noninterference}
& $\proj{\vec{t}_1}_L$ $\pref$
$\proj{\vec{t_2}}_L$   &
$\closure{m}_{L,\Gamma} $  & = \\  \hline
\fi
\leftcentering{Gradual Release}
& $\proj{\vec{t}_1}_L$ $\pref$
$\proj{\vec{t_2}}_L$     &
$ \closure{m}_{L,\vec{t}^{[i]}.\gamma} \cap \mathcal{K}(c,\vec{t}^{[:i-1]}, 
\sim_{\cod{GR}}) $   & =\\ \hline
%
\iftightgr
\leftcentering{Tight Gradual Release}
& $\proj{\vec{t}_1}_L$ $\pref$
$\proj{\vec{t_2}}_L$    &
$ \closure{m}_{L,\vec{t}^{[i]}.\gamma} $  & = \\ \hline
\fi
\leftcentering{According to Policy} 
& \small{$\exists R. ~\forall (i, j)\in R.~
$}  $\proj{\vec{t_1}^{[i]}}_{b,L}
\cong 
\proj{\vec{t_2}^{[j]}}_{b,L}$ 
&
$\closure{m}_{\not =b} $  & =\\ \hline
\leftcentering{Cryptographic Erasure}
& $\proj{\vec{t_1}}_L =  \proj{\vec{t_2}}_L^{[i:j]} $ &
$\bigcap_{t \in \vec{t} }\closure{m}_{L,t.\gamma} $
 & = \\\hline
\ifforgetful
\leftcentering{Forgetful Attacker}
& $\exists \vec{t}' \pref \vec{t}_2.~\cod{atk}(\proj{\vec{t}_1}_L) = 
\cod{atk}(\proj{\vec{t}'}_L)$ 	&
$ \closure{m}_{L, \vec{t}^{[i]}.\gamma}  \cap \mathcal{K}(c, 
\vec{t}^{[:i-1]}, 
\sim_{\cod{FA}})  
$ & =  \\\hline
\fi
\leftcentering{Paralock}
& $\proj{\vec{t}_1}_A$ $\pref$
$\proj{\vec{t_2}}_A$     &
$ \begin{cases} 
\closure{m}_A \cap \mathcal{K}(c, \vec{t}^{[:{i-1}]},\sim_{\cod{PL}}), & 
\vec{t}_{[i]}.\Delta  \subseteq \lockset_A\\
\closure{m}_\emptyset, &  \text{otherwise}
\end{cases}$  & =\\ \hline
\leftcentering{\policyname}
& $\proj{\vec{t}_1}_L$ $\pref$
$\proj{\vec{t_2}}_L$ 
 &
$ \begin{cases} 
\closure{m}_{\not =b}, & b \text{ is transient} \\
\closure{m}_{\not =b} \cap \mathcal{K}(c, \vec{t}^{[:{i-1}]},\sim), & b \text{ 
is persistent}
\end{cases}$
& $\proj{\vec{t}_1}_{b,L}^{[n]} = \proj{\vec{t}_2}_{b,L}^{[n]}$
\\\hline
\end{tabular}
}
\caption{Existing End-to-End Security Policies and \policyname{}
Written in the Formalization Framework.}
\label{tab:summary}
\end{table*}

\subsection{Formalization Framework for Dynamic Policies}

We first abstract way a few building blocks of Definition~\ref{def:newpolicy}.
To define them more concretely, we consider an output sequence $\vec{t}$
produced by $\configs{c,m}$, i.e., $\configs{c, m} \termout \vec{t}$, as the
context.

As already discussed in Section~\ref{sec:semantics}, the building blocks are:
\begin{itemize}
\item \emph{Output Indistinguishability, written as $\sim$}: two output
sequences $\vec{t_1}$ and $\vec{t_2}$ satisfies $\vec{t_1}\sim \vec{t_2}$
when they are considered indistinguishable to the attacker.

\item \emph{Policy Allowance, written as $\allowence(m,\vec{t},b,L)$}: a set of
initial memory that should be indistinguishable to attacker at $L$ at the end
of sequence $\vec{t}$.

\item \emph{Consistency Relation, written as $\equiv$}: when trying to
precisely define the knowledge gained from each output event, two sequences are
considered ``consistent'', even if they are not identical
(Definition~\ref{def:consistency}).

\end{itemize}

With the abstracted parameters, we first generalize the knowledge definition of
$k_1$ (Equation~\ref{eq:knowledge}) on an arbitrary relation $\sim$ on output
sequences:

%



\begin{Definition}[Generalized Knowledge]
\label{eq:genknowledge}
	\begin{equation}
	\label{eq:kout}
	\mathcal{K}(c,\vec{t}, \sim ) \defn \{ m~\mid~
	\configs{c, m} \termout \vec{t'} \AND \vec{t} \sim \vec{t'}~
	\} 
	\end{equation}
	
\end{Definition}

Therefore, with abstract $\sim$, $\allowence(m,\vec{t},b,L)$ and $\equiv$,
we can generalize Definition~\ref{def:newpolicy} as the following framework:
\begin{Definition}[Formalization Framework]
	\label{def:framework}
	Given trace indistinguishability relation $\sim$, \consistency ~relation 
	$\equiv$
	and policy allowance $\allowence$, a command $c$ satisfies a dynamic policy
	iff the knowledge gained from observing any output does not exceed its corresponding
	policy allowance:
	\begin{multline*}
	\forall m, L\subseteq \mathbb{L}, b\in \mathbb{B}, \vec{t}.~\configs{c,
		m} \termout \vec{t}  
	\implies 
	\forall 1\leq i\leq \len{\vec{t}}.~\\
	\bigcup_{\exists m',j.~\configs{c, m'} \termout \vec{t} \land 
	\vec{t'}^{[:j]} \equiv \vec{t}^{[:i]} }
	~\mathcal{K}(c,\vec{t'}^{[:j]}, \sim) \supseteq 
	\allowence(m,\vec{t}^{[:i]},b,L)
	\end{multline*}
\end{Definition}

Let $\sim_{\cod{DR}} \defn \{(\vec{t}_1,\vec{t}_2)~|~\proj{\vec{t}_1}_L \pref
oump\proj{\vec{t}_2}_L\}  $, $\allowence_{\cod{DR}}$ be as defined in
\text{Equation~(\ref{eq:allowance})}, and $\equiv_{\cod{DR}}$ be as defined in
Definition~\ref{def:eff-knowledge}, it is easy to check that
Definition~\ref{def:framework} is instantiated to
Definition~\ref{def:newpolicy}.

Moreover, when $\equiv$ is instantiated with an equality relation $=$, a case
that we have seen in all existing dynamic policies, the general framework can
be simplified to the following form:
\begin{multline*}
	\forall c, m, L\subseteq \mathbb{L}, b\in \mathbb{B}, \vec{t}.~\configs{c,
		m} \termout \vec{t}  
	\implies 
	\forall 1\leq i\leq \len{\vec{t}}.~\\
	\mathcal{K}(c,\vec{t}^{[:i]}, \sim) \supseteq 
	\allowence(m,\vec{t}^{[:i]},b,L)
\end{multline*}

We use this simpler form for any dynamic policy where consistency is
simply defined as equivalence.

\subsection{Existing works in the formalization framework}
\label{sec:encoding}
Next, we incorporate existing definitions into the formalization framework; the
results are summarized in Table~\ref{tab:summary}. We first highlight a few
insights from Table~\ref{tab:summary}. \revise{Then, for each work (except for
Paralock due to space constraint), we 
sketch how to convert it (with potentially different security specification
language and semantic formalization) into the specification language in
Figure~\ref{fig:while-syntax} and Definition~\ref{def:framework} respectively.
The conversion of Paralock and the correctness proofs of all conversions are
available in the Supplementary Material.}


\subsubsection{Insights from Table~\ref{tab:summary}}

To the best of our knowledge, this is the first work that enables
apple-to-apple comparison between various dynamic policies. We highlight a few
insights.

First, an erasure policy (e.g., According to Policy and Cryptographic Erasure)
defines indistinguishability $\sim$ in a substantially more complicated way
compared with others. The complexity suggests that formalizing an erasure
policy is more involved compared with other dynamic policies.

Second, besides \policyname{}, Gradual Release, Paralock and Forgetful Attacker
also have $ \mathcal{K}(c,\vec{t}^{[:i-1]}, \sim) $ as part of policy
allowance. Recall that $ \mathcal{K}(c,\vec{t}^{[:i-1]}, \sim) $ represents the
\emph{past knowledge} excluding the last output on $\vec{t}$.  Hence, these
policies are \emph{persistent} policies. On the other hand,  all other dynamic
policies are transient policies.

Third, since an erasure policy by definition is transient, persistent policies
such as Gradual Release and Paralock cannot check erasure policy, such as the
example in Figure~\ref{fig:application}-B: leaking credit card after erasure
violates the erasure policy.

\subsubsection{Gradual Release}
\label{sec:gradualrelease}
Gradual Release assumes a mapping $\Gamma$ from variables to levels in a
Denning-style lattice.  A release event is generated by a special command $
x:=\cod{declassify}(e) $. Informally, a program is secure when illegal flow
w.r.t. $\Gamma$ only occurs along with release events. 
Hence, we encode a release event as
\[\cod{EventOn}(r);x:=e;\outcmd{\Gamma(x)}{e};\cod{EventOff}(r);\] where $r$ is
a distinguished event for release, and we set
$\forall x. ~\Gamma'(x)=r?\mathbb{L}\leftrightarrows\Gamma(x)$ to state that
any leakage of any variable is allowed when this is a release event, but 
otherwise, the information flow restriction of $\Gamma$ is obeyed.

Gradual Release is formalized on the insight that ``knowledge must remain
constant between releases'':

\begin{Definition}[Gradual Release~\cite{askarov2007}]
	\label{def:gradualrelease}
	A command $c$ satisfies gradual release w.r.t. $\Gamma$
if\footnote{\revise{Note that $\configs{c, m} \termout \vec{t}$ only considers
terminating program executions by definition. So we used the
termination-insensitive version of Gradual Release.}}
	\begin{multline*}
	\forall c, m, L, i, \vec{t}.~\configs{c, m} \termout \vec{t}  
	\implies 
	\\
	\forall $i$ \text{ not release event}.~k(c,m,\vec{t}^{[:i]},L,\Gamma) 
	= k(c,m,\vec{t}^{[:i-1]},L,\Gamma) 
	\end{multline*}
	where $k(c,m,\vec{t},L,\Gamma) \defn $
	\begin{equation}
	\label{def:k_GR}
	\{ m'\mid m'\in \closure{m}_{L,\Gamma}  
	\AND 
	~\configs{c, m} \termout \vec{t'} \AND 
	\vec{t} \pref \vec{t}'\}
	\end{equation}

\end{Definition}

While the original definition does not immediately fit our
framework, 
we prove that they are equivalent by:
\[\sim_{\cod{GR}} \defn \{(\vec{t}_1,\vec{t}_2)~|~\proj{\vec{t}_1}_L \pref
\proj{\vec{t}_2}_L\}  \quad \equiv_{\cod{GR}}\defn =
\]
\[\allowence_{\cod{GR}} \defn 
\closure{m}_{L,~ \vec{t}^{\len{\vec{t}}}.\gamma} \cap \mathcal{K}(c, \vec{t}^{[:
	\len{\vec{t}}-1]}, \sim_{\cod{GR}})  \]

Recall that in our encoding, 
a release event emits an output event $ \configs{\Gamma(x), e, \gamma_{\bot}}
$, where $ \gamma_{\bot} $ maps all variable to public. This
essentially makes the allowance check $\mathcal{K}(\dots) \supseteq \allowence$
trivially true, resembling Definition~\ref{def:gradualrelease}.

\begin{Lemma}
	\label{lemma:gradualrelease-eq}
	With $\sim \defn \sim_{\cod{GR}}$, $\equiv\defn \equiv_{\cod{GR}}$ and
	$\allowence \defn \allowence_{\cod{GR}}$, Definition~\ref{def:framework} is
	equivalent to Definition~\ref{def:gradualrelease}.
\end{Lemma}

\textbf{Observation.}
From Table~\ref{tab:summary}, it is obvious that Gradual Release uses
indistinguishability $\sim_L$. Its policy allowance is defined by the last
dynamic specification $ \vec{t}^{[\len{\vec{t}}]}.\gamma $, as well as the
knowledge gained from previous outputs.

\subsubsection{Tight Gradual Release} 
Tight Gradual Release~\cite{Askarov:2009csf,askarov2007local} is an extension
of Gradual Release. Similar to Gradual Release, it assumes a base policy $
\Gamma $ and uses a $x:=\cod{declassify}(e) $ command to declassify the value of
$ e $. However, the encoding of declassification command is different for two
reasons. First, we can only encode a subset of Tight Gradual Release where
declassification command contains $ \cod{declassify}(x) $, since our
language does not fully support partial release (Section~\ref{sec:dimensions}).
Second, declassification in Tight Gradual Release is both precise (i.e., only
variable $x$ in $ \cod{declassify}(x) $ is downgraded) and permanent (i.e., the
sensitivity of $x$ cannot upgrade after $x$ is declassified). Hence, we encode
$ x' :=\cod{declassify}(x) $ as
\[\cod{EventOn}(r_x);x':=x;\outcmd{\Gamma(x')}{x};\] 
where $r_x$ is a distinguished security event for releasing just $x$, and we
set $ \Gamma'(x)=r_x?\mathbb{L}\leftarrow\Gamma(x)$ to state that $x$ is
declassified once $r_x$ is set.

Tight Gradual Release uses the same knowledge definition from Gradual Release,
but its execution traces also dynamically track the set of declassified variables
$X$:
\[
\configs{c, m, \emptyset} \rightarrow ^* \configs{c', m', X}
\]

\begin{Definition}[Tight Gradual Release] A program c is secure if for any  
	trace $ \vec{t} $, initial memory $ m $ and attacker at level $ L $, we have
	\begin{align*}
	\forall i. ~1 \leq i \leq \len{\vec{t}}.~  (\closure{m}_{L, \Gamma} \cap 
	\closure{m}_{X_i} )
	\subseteq k(c,m, \vec{t}^{[:i]}, L, \Gamma)
	\end{align*}
	\label{def:tightGR}
	where $ X_i $ is the set of declassified variables associated with the 
	$ i $-th output.
\end{Definition}

Due to the encoding of declassification commands, we know that for each output
at index $i$ in $ \vec{t} $ we have:
\[
\closure{m}_{L, \vec{t}^{\len{\vec{t}}}.\gamma} = (\closure{m}_{L, \Gamma} \cap 
\closure{m}_{X_i} )\]

Hence, we can rephrase Tight Gradual Release as follows:
\[ \sim_{\cod{TGR}} 
\defn
\{(\vec{t_1},\vec{t_2})\mid \proj{\vec{t_1}}_L \pref
\proj{\vec{t_2}}_L  \} \]
\[\equiv_{\cod{TGR}} \defn = ~~~~~~ \allowence_{\cod{TGR}} \defn 
\closure{m}_{L, \vec{t}^{\len{\vec{t}}}.\gamma}  \]

\begin{Lemma}
	\label{lemma:tightGR-eq}
	With $\sim \defn \sim_{\cod{TGR}}$, $\equiv\defn \equiv_{\cod{TGR}}$ and
	$\allowence \defn\allowence_{\cod{TGR}} $, 
	Definition~\ref{def:framework} is
	equivalent to Definition~\ref{def:tightGR}.
\end{Lemma}

\textbf{Observation:}
%
Tight Gradual Release is more precise than Gradual Release since the encoding
of $\cod{declassify}(x)$ precisely downgrades the sensitivity of $x$ but not
any other variables, while the encoding for Gradual Release downgrades
all variables.

\revise{
Compared to \policyname, the most important difference is that the consistency
relation $\equiv$ is defined in completely different ways.  As discussed in
Section~\ref{sec:challenges}, it is important to define it properly for general 
dynamic policies. 
The other major difference is that the security semantics of
Tight Gradual Release cannot model erasure policies.
Consider the example in
Figure~\ref{fig:application}-B.i with $m_1(credit\_card)=0$,
$m_2(credit\_card)=1$ and attacker level $M$. Given a program execution
$\configs{c, m_1} \termout \vec{t}$, we have $\mathcal{K}(c, \vec{t}^{[:i]},
\sim_{\cod{TGR})}=\{m_1\}$ for all $i\geq 1$. However, $ credit\_card $ is
upgraded from $ M $ to $ \top $ when $i=2$ (i.e., the second output), the
secure program (i) is incorrectly rejected since $\mathcal{K}(c,
\vec{t}^{[:2]}, \sim_{\cod{TGR}})= \{m_1\}\not\supseteq \{m_1,m_2\} =
\closure{m_1}_{M,\vec{t}^{[2}.\gamma}$. 
}

\subsubsection{According to Policy} Chong and Myers propose noninterference
according to policy \cite{chong2005,chong2008} to integrate erasure and
declassification policies. We use the formalization in the more recent
paper~\cite{chong2008} as the security definition.

This work uses \emph{compound labels}, a similar security specification as
ours: a label is is either a simple level $ \ell $ drawn from a Denning-style
lattice, or in the form of $ q_1 \abv{e}{\rightarrow}  q_2 $, where $q_1$ and
$q_2$ are themselves compound labels. Hence, converting the specification to
ours is straightforward.

Noninterference according to policy is defined for each variable in a two-run
style. In particular, it requires that for any two program executions where the
initial memories differ \emph{only} in the value of the variable of interest,
their traces are indistinguishable regarding a \emph{correspondence} $R$:
\begin{Definition}[Noninterference According To Policy~\cite{chong2008}] 
	\label{def:niaccord}
	A program $ c $ is noninterference according to policy if for any variable 
	$x$ 
	(with
	policy $b$) we have:\footnote{The original definition uses a specialized 
		\emph{label
			semantics}, denoted as $ \llbracket b \rrbracket_{\configs{c,m}} $, 
			and
		requires $(\configs{c_i,m_i},\ell)\not\in \llbracket b
		\rrbracket_{\configs{c,m}}$ which means that if by the time 
		$\configs{c,m}$
		reaches state $\configs{c_i,m_i}$, confidentiality level $\ell'$ may not
		observe the information. It is easy to convert that to $\ell\not\in
		\psemantics{b}{ \trace_{[:i]} }$ in our notation.}
	\begin{multline*}
	\forall m_1, m_2, \ell, \vec{t}_1, \vec{t}_2.~\forall y\not= 
	x.~m_1(y)=m_2(y)\\
	\AND \configs{c, m_1} \termout \vec{t_1} 
	\AND \configs{c, m_2} \termout \vec{t_2} \implies \\ 
	\exists R.~~ 
	\Big(
	\forall (i,j)\in R, \ell.~
	\ell \not \in \llbracket b \rrbracket_{{\trace_1}_{ [:i]}} 
	\AND \ell \not \in \llbracket b \rrbracket_{ {\trace_2}_{[:j] }  } 
	\sat 
	{\trace_{[i]} \loweq_{\ell} \trace'_{[j]}}
	\Big)
	\end{multline*}
\end{Definition}
where a \emph{correspondence} $R$ between traces $\trace_1$ and $\trace_2$ is a
subset of $\nat × \nat$ such that:
\begin{enumerate}
	\item (Completeness) either $\{i~\mid~(i,j)\in R\} = \{i\in 
	\nat~\mid~i<|\trace_1|\}$ or
	$\{j~\mid~(i,j)\in R\} = \{j\in \nat~\mid~j<|\trace_2|\} $, and
	
	\item (Initial configurations) if $\len{R}>0$ then $(0,0)\in R$, and
	
	\item (Monotonicity) for all $(i,j)\in R$ and $(i',j')\in R$, if $i<i'$ then
	$j\leq j'$ and symmetrically, if $j<j'$ then $i\leq i'$.
\end{enumerate}

To transform Definition~\ref{def:niaccord} to our framework, we make a few
important observations:
\begin{itemize}
	\item The definition relates two memories that differ in \emph{exactly one
		variable} (i.e., $\forall y\not= x.~m_1(y)=m_2(y)$), which is different 
		from
	the usual low-equivalence requirement in other definitions. However, it is 
	easy
	to prove that (shown shortly) it is equivalent to a per-policy definition
	$ \closure{m}_{\not = b} $ 
	in our framework, that considers memories that differ \emph{only} for 
	variables
	with a particular policy $b$.

	\item The component of $\ell \not \in \llbracket q 
	\rrbracket_{\vec{t_1}^{[:i]}} \AND \ell \not \in \llbracket q 
	\rrbracket_{\vec{t_2}^{[:j]}}$ filters out
	non-interesting outputs, which functions the same as the filtering function 
	$  \proj{\vec{t}}_{b,L}$. 
	\item We define $ \cong $ on two output sequence as below:
	\[ \vec{t_1} \cong 
	\vec{t_2} \iff \neg (\len{\vec{t_1}} = \len{\vec{t_2}} \AND \exists i.~ 
	\vec{t}_1^{~[i]} \not = \vec{t}_2^{~[i]} ) \]
\end{itemize}

Based on the observations, we convert Definition~\ref{def:niaccord} into our
framework as follows:
\[
\sim_{\cod{AP}} \defn \{ (\vec{t_1},\vec{t_2})\mid \exists R. ~\forall (i, 
j)\in 
R.~\proj{\vec{t_1}^{[i]}}_{b,L}
\cong 
\proj{\vec{t_2}^{[j]}}_{b,L}  \}
\]
\[
\equiv_{\cod{AP}} \defn = ~~~~~~ 
\allowence_{\cod{AP}} \defn \closure{m}_{\not =b}
\]

\begin{Lemma}
	\label{lemma:according-eq}
	With $\sim \defn \sim_{\cod{AP}}$, $\allowence \defn
	\allowence_{\cod{AP}}$, and outside equivalence $ \equiv \defn  
	\equiv'_{\cod{AP}}$, Definition~\ref{def:framework} is equivalent to
	Definition~\ref{def:niaccord}.
\end{Lemma}
\textbf{Observation:} Compared with Gradual Release and Tight Gradual
Release, the most interesting component of According to Policy is in its unique
indistinguishability definition, which uses the correspondent relationship $ R
$. Intuitively, According to Policy relaxes the indistinguishability definition
in the way that two executions are indistinguishable as long as a
correspondence $ R $ exists to allow decreasing knowledge. However, as shown
later in the evaluation, the relaxation with $ R $ could be too loose: it
falsely accepts insecure programs.

\subsubsection{Cryptographic Erasure}
Cryptographic erasure~\cite{askarov2015} uses the same compound labels to 
describe erasure policy and knowledge is defined as:
\begin{multline*}
k_{\cod{CE}}(c,L, \vec{t})=\{ m~|~\configs{c, m} \xrightarrow{\vec{t_1}}{^*} 
\configs{c_1, m_1} \xrightarrow{\vec{t_2}}{^*}  \configs{c', m'} \\
\AND \proj{\vec{t_2}}_{L} = \proj{\vec{t}}_{L} \} 
\end{multline*}
Unlike other policies, the definition specifies knowledge
based on the \emph{subtrace} relation, rather than the standard \emph{prefix}
relation. The reason is that it has a different attack model: it assumes an
attacker who might \emph{not} be able to observe program execution from the
beginning.

\begin{Definition}[Cryptographic Erasure Security~\cite{askarov2015}]
	\label{def:crypto}
	A program $ c $ is secure if any execution starting with memory $ m $, the 
	following holds:
	\begin{multline*}
	\forall c_0, m_0, c_i,m_i, c_n, m_n,
	\vec{t_1}, 
	\vec{t_2}, 
	L, i, n.~~\\
	\configs{c_0,m_0} 
	\xrightarrow{\vec{t_1}}{^*} \configs{c_i, m_i} 
	\xrightarrow{\vec{t_2}}{^*}  \configs{c_n, m_n}\\
	\sat k_{\cod{CE}}(c, L, \vec{t_2}) \supseteq \bigcap_{t \in 
		\vec{t}_{2} }\closure{m}_{L, t.\gamma}
	\end{multline*}
\end{Definition}
To model subtraces, we adjust the $\forall 1\leq i \leq \len{\vec{t}}$
quantifier in the framework with $\forall 1\leq i < j \leq
\len{\vec{t}}$, and write $\vec{t}[i:j]$ for the subtrace between $i$ and $j$.
Then, converting Definition~\ref{def:crypto} into our framework is relatively
straightforward: 
\[ \sim_{\cod{CE}} 
\defn
\{(\vec{t_1},\vec{t_2})\mid \proj{\vec{t_1}}_L \text{ subtrace of } 
\proj{\vec{t_2}}_L  \} \]
\[\equiv_{\cod{CE}} \defn = ~~~~~~ \allowence_{\cod{CE}} \defn \bigcap_{t \in 
	\vec{t} }\closure{m}_{L,t.\gamma}  \]

\begin{Lemma}
	\label{lemma:crypto-eq}
	With $\sim \defn \sim_{\cod{CE}}$, $\equiv\defn \equiv_{\cod{CE}}$ and
	$\allowence \defn \allowence_{\cod{CE}}$, Definition~\ref{def:framework} 
	with
	adjusted attack model is equivalent to Definition~\ref{def:crypto}.
\end{Lemma}

\textbf{Observation:} 
Compare with other works, the most interesting part of cryptographic erasure is
that its indistinguishability and policy allowance are both defined on
subtraces; moreover, the latter uses the weakest policy on the subtrace.
Intuitively, we can interpret Cryptographic Erasure security as: the
subtrace-based knowledge gained from observing a subtrace should be bounded by
the smallest allowance (i.e, the weakest policy) on the trace.  

\subsubsection{Forgetful Attacker}
Forgetful Attacker~\cite{askarov2012, Delft:2015post} is an expressive policy
where an attacker can ``forget'' some learned knowledge. To do so, an attacker
is formalized as an automaton $ \cod{Atk}\configs{Q_A, q_{init},
\delta_A} $, where $ Q_A $ is a set of attacker's states, $ q_{init} \in Q_A $
is the initial state, and $ \delta_A $ is the transition function. 
The attacker observes a set of events produced by a program execution, and
updates its state accordingly:
\begin{align*}
\cod{Atk}(\epsilon)=& q_{init}\\
\cod{Atk}(\vec{t}_{\pref i}) =& \delta(\cod{Atk}_A(\vec{t}_{\pref i-1}), t^{[i]})
\end{align*} 

Given a program $c$, an automaton $ \cod{Atk} $ and attacker's level $ L $,
knowledge is defined as the set of initial memory that could have resulted in
the same state in the automaton:
\begin{multline*}
k_{\cod{FA}}(c, L, \cod{Atk}, \vec{t} ) = \{m~|~ \configs{c,m} 
\xrightarrow{\vec{t_1}}{^*} 
\configs{c',m'} \xrightarrow{\vec{t_2}}{^*}  m'' \\
\AND \cod{Atk}(\proj{\vec{t_1}}_{L}) = 
\cod{Atk}(\proj{\vec{t}}_{L}) \}
\end{multline*}

\begin{Definition}[Security for Forgetful Attacker~\cite{askarov2012}]
\label{def:fa_original}
	A program $ c $ is secure against an attacker $ \cod{Atk}\configs{Q_A, 
		q_{init}, 
		\delta_A)} $ with level $ L $ if:
	\label{def:forgetful}
	\begin{multline*}
	\forall c, c', m, m', \vec{t}, t', L.~
	\configs{c, m_1} \termout \vec{t} \cdot t' \sat 
	\\
	k_{\cod{FA}}(c, L, \cod{Atk}, \vec{t} \cdot t') ~\supseteq~ 
	k_{\cod{FA}}(c, L, \cod{Atk}, \vec{t}) ~\cap~ 
	\closure{m}_{L , \gamma'}	
	\end{multline*}
\end{Definition}


The conversion of Definition~\ref{def:forgetful} to our framework is
straightforward:
\[ \sim_{\cod{FA}} 
\defn
\{(\vec{t_1},\vec{t_2})\mid \exists \vec{t'} \pref 
\vec{t_2}.~\cod{Atk}(\vec{t_1}) 
= \cod{Atk}(\vec{t'})\} \]
\[\equiv_{\cod{FA}} \defn = ~~~~~~ \allowence_{\cod{FA}} \defn\mathcal{K}(c,
\vec{t}^{[:\len{\vec{t}}-1]}, \sim_{\cod{FA}})
~\cap~ \closure{m}_{L,\vec{t}^{[\len{\vec{t}}]}.\gamma } \]

\begin{Lemma}
	\label{lemma:forgetful-eq}
	With $\sim \defn \sim_{\cod{FA}}$, 
	$\allowence \defn \allowence_{\cod{FA}}  $, and outside equivalence
        $\equiv\defn \equiv_{\cod{FA}}$,
	Definition~\ref{def:framework} is equivalent to 
	Definition~\ref{def:forgetful}.
\end{Lemma}

\revise{
\textbf{Observation:} We note that Forgetful Attacker
(Definition~\ref{def:fa_original}) was originally formalized in the same format
as \policyname{} (the persistent case). However, there are various differences in
the modeling, as can be observed from Table~\ref{tab:summary}. Most
importantly, Forgetful Attacker security is parameterized by an automaton
$\cod{Atk}$; in other words, a program might be both ``secure'' and
``insecure'' depending on the given automaton.  Consider the program in
Figure~\ref{fig:application}-B(i). The program satisfies Forgetful Attacker
security with any automation that forgets about the credit card information.
Nevertheless, characterizing such “willfully stupid” attackers is an open
question~\cite{askarov2012}. Second, the definition of the consistency relation
$\equiv$ is completely different.  As discussed in
Section~\ref{sec:challenges}, it is important to define it properly to allow
information flow restrictions to downgrade and upgrade in arbitrary ways.}
%

\section{Evaluation}
\label{sec:evaluation}


\newcommand{\tablength}{0.45}
\newcommand{\facetlength}{0.45}
\newcommand{\statlength}{0.3}
\begin{table*}
	\centering
\scalebox{1}{
\begin{tabular}{|l||C{\tablength cm}C{\tablength cm}|
		C{\tablength cm}C{\tablength cm}|C{\tablength cm}
		C{\tablength cm} ||
		C{\statlength cm}| C{\statlength cm} | C{\statlength cm} |
		C{\statlength cm}| C{\statlength cm} | C{\statlength cm} || }
\hline
& \multicolumn{6}{C{4cm}|}{{Examples in Fig~\ref{fig:application}}} 
& \multicolumn{3}{C{1.8cm}|}{{Existing(35)}} 
& \multicolumn{3}{C{1.8cm}||}{{New (23)}} 
\\ \cline{2-13}
& A(i) & A(ii) & B(i) & B(ii) & C(i) & C(ii) 
& \yes & \no &\na & \yes & \no &\na  \\  
\hline
{Gradual Release} 
&\yes&\yes&\na&\na&\yes& \yes 
&28&2&5&14&1&8 \\
Tight Gradual Release 
&\yes&\yes&\na&\na&\yes& \yes 
&18&0&17&8&0&15 \\
According to Policy p
&\yes &\yes&\yes &\no&\na&\na 
&17&6&12&12&4&7 \\	
Cryptographic Erasure 
&\na&\na&\yes&\yes&\na&\na 
&21&0&14&7&1&15\\ 
\ifforgetful
{Forgetful Attacker-Single} 	
&\yes&\yes&\yes&\no&\yes&\yes 
&31&4&0&19&4&0\\	
\fi
\hline
\policyname &\yes&\yes&\yes&\yes&\yes&\yes 
&35&0&0&23&0&0 \\
	\hline
\end{tabular}
}

\vspace{0.3em}
\footnotesize{ 
	`\yes '  means the policy checks the program as intended (same as ground 
	truth);  `\no' means the policy fails to check \\the program as intended.
	~`-'~ means the program is not in the scope of the policy (not applicable).
}
\caption{Evaluation Results.} 
\vspace{-3ex}
\label{tab:result}
\end{table*}

In this section, we introduce  \anntrace~benchmark and implement the
dynamic policies as the form shown in Table~\ref{tab:summary}. The benchmark
and implementations are available on 
github\footnote{\url{https://github.com/psuplus/AnnTrace}}.


\subsection{\anntrace{} Benchmark}
To facilite testing and understanding of dynamic policies, we created the
\anntrace{} benchmark. It consists of a set of programs annotated with
\emph{trace-level} security specifications. Among 58 programs in the benchmark,
35 of them are collected from existing
works~\cite{askarov2007, askarov2012, askarov2015,
sabelfeld2003, chong2008, broberg2009}.
References to the original examples are annotated in the benchmark programs.
The benchmark also includes 23 programs that we created, such as the programs 
in Figure~\ref{fig:application}, and the counterexamples in 
Figure~\ref{fig:examples}.
\begin{figure}
	\centering
	\includegraphics[clip, trim=0cm 0cm 0cm 0cm, 
	width=0.4\textwidth]{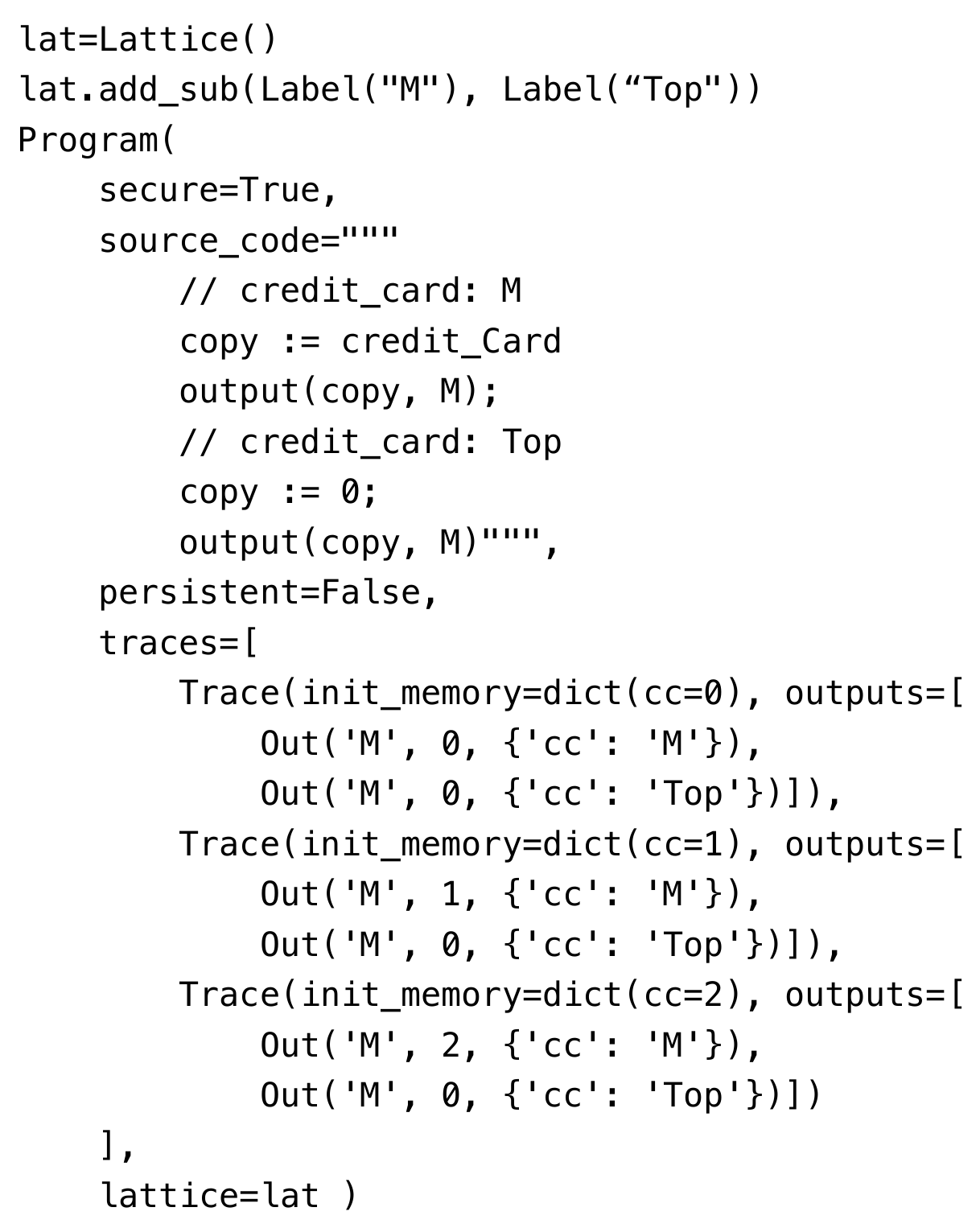}
	\caption{Annotated Program for Fig.~\ref{fig:application}-B(i)}
	\label{fig:annotations}
\end{figure}

The benchmark is written in Python. Fig.~\ref{fig:annotations} shows an example
of annotated program for the source code in Fig.~\ref{fig:application}-B(i). As
shown in the example, each program 
consists of:
\begin{itemize}
\item \textit{secure}, a boolean value indicating whether this program is a 
secure program; the ground truth of our evaluation.
\item \textit{source code},  written in the syntax shown in
Fig~\ref{fig:while-syntax};
\item \textit{persistent}, a boolean value indicating whether the intended policy
in this program is persistent (or transient);
\item \textit{lattice}, $ \Lattice $ , the security lattice used by the 
program\footnote{We use lattice instead of level set for conciseness in the
implementation.}; 
\item \textit{traces}, executions of the program. Each trace $ \trace 
$ has:
\begin{itemize}
	\item \textit{initial memory}, $ m $ , mapping from variables to integers
	\item \textit{outputs}, $ \vec{t} $, a list of output events, each $ 
	t $ in type $ \cod{Out} $:
	\begin{itemize}
		\item \textit{output level}, $ \ell $ , a level from the lattice $ 
		\Lattice $
		\item \textit{output value}, $ v $, an integer value
		\item \textit{policy state}, $ \gamma $, mapping from variables to 
		levels
	\end{itemize}
\end{itemize}
\end{itemize} 

Given a program in existing work, we (1) use the claimed security of code as
the ground truth, (2) convert the program into our specification language and to a
security lattice, (3) mark persistent (or transient) according to if the
correponding paper presents a persistent (or transient) policy, and (4)
manually write down a finite number of traces that are sufficient for checking
the dynamic policy involved in the example.  

\subsection{Implementation}
We implemented all dynamic policies in Table~\ref{tab:summary} in Python,
according to the formalization presented in the table. With exception of
Forgetful Attacker and Paralocks, all implemented policies can directly work on
the trace annotation provided by the \anntrace~benchmark. Forgetful Attack
policy requires an automaton as input. So we use a single memory automaton that
only remembers the last output and forgets all previous outputs. Paralocks
security requires ``locks'' in a test program but most tests do not have locks. So
we are unable to directly evaluate it on the
\anntrace~benchmark.\footnote{Although we are unable to evaluation Paralocks
directly, we believe its results should resemble those of Gradual Release, as
its security condition is a generalisation of the gradual release
definition~\cite{Broberg:2010popl}.}

Existing policies are not generally applicable to all tests. Recall that each
test has a persistent/transient field. Moreover, for each test, we
automatically generate the following two features from the \emph{traces} field:
\begin{enumerate}
\item[A.] there is no policy upgrading in the trace;
\item[B.] there is no policy downgrading in the trace;
\end{enumerate} 

These tags are used to determine if a concrete policy is appliable to the test.
For example, Cryptographic Erasure is a transient policy that only allows
upgrading. Hence, it is applicable to the tests with tag \emph{transient} and
B.

\subsection{Results}

The evaluation results are summerized in Table \ref{tab:result}. 
For the examples shown in Figure~\ref{fig:application} (classical examples for
declassification, erasure and delegation/revocation), we note that
\policyname{} is the only one that is both applicable and correct in all cases.

Among the 35 programs collected from prior papers and the 23 new programs,
\policyname{} is still both applicable and correct to all programs.
In contrast, the existing works fall short in one way or another: with limited
applicability or incorrect judgement on secure/insecure programs.
%
%
%
%
Interestingly, According to Policy, Cryptographic Erasure
and Gradual Release all make wrong judgment on some corner cases. Here, we
discuss a few representative ones.

For According to Policy, the problematic part is the $ R $ relation. The policy
states that as long as a qualified $ R $ can be found to satisfy the equation,
a program is secure. We found that the restriction on $ R $ is too weak in many
cases: a qualified $ R $ exists for a few insecure programs.

For Crypto-Erasure policy, the failed examples is shown in
Figure~\ref{fig:examples}-(A). It is an insecure program as the attacker learns
that $ x = 0 $ if two outputs are observed. However, 
Crypto-Erasure accepts this program as secure for the reason that their policy
ignores the location of an output. In this example, for the output $ 0 $, the
security definition of Crypto-Erasure assumes that two executions are indistinguishable to the
attacker if there exists a $ 0 $ output anywhere in the execution. Therefore,
an execution with a single $ 0 $ output appears indistinguishably to the
execution with two $ 0 $ outputs (both exists a $ 0 $ output). Thus, the policy
fails to reject this program.

\newsavebox\FilteringExample
\begin{lrbox}{\FilteringExample}
\begin{minipage}{0.25\textwidth}
\begin{lstlisting}[numbers=none,xleftmargin=5 
pt,framexleftmargin=15pt]
// h, h1: {D}$\Rightarrow$a
// l, l2: {}$\Rightarrow$a
$ \cod{open} $(D);
$ \cod{if} $ (h) { l2:=h1;}  
$ \cod{close} $(D);
$$l:=0; 
\end{lstlisting}
\end{minipage}
\end{lrbox}
	
\newsavebox\FilteringMemExample
\begin{lrbox}{\FilteringMemExample}
	\begin{minipage}{0.17\textwidth}
\begin{lstlisting}[numbers=none,xleftmargin=5 
pt,framexleftmargin=5pt]
// x : L
$\cod{output}$($ L $, 0);
// x : H
if (x == 0)
  $\cod{output}$($ L $, 0);
		\end{lstlisting}
	\end{minipage}
\end{lrbox}
		
\newsavebox\AlignExampleA
\begin{lrbox}{\AlignExampleA}
\begin{minipage}{\boxwidth\textwidth}
\begin{lstlisting}[numbers=none,xleftmargin=\leftmarginvalue 
pt,framexleftmargin=15pt]
// h, h2: H,
if (h) {
   // h2: L
   $\cod{output}$($ L $, h2);
}
// h, h2: H,
$\cod{output}$($ L $, 0);
\end{lstlisting}
\end{minipage}
\end{lrbox}

\newsavebox\AlignExampleB
\begin{lrbox}{\AlignExampleB}
\begin{minipage}{\boxwidth\textwidth}
\begin{lstlisting}[numbers=left,xleftmargin=\leftmarginvalue 
pt,framexleftmargin=15pt]
// h, h2: H,
if (h) {
	// h2: L
	$\cod{output}$($ L $, h2);
	$\cod{output}$($ L $, 0);
}
\end{lstlisting}
\end{minipage}
\end{lrbox}

\begin{figure}
	\centering
	\begin{tabular}{|c|c|}
		\hline
		 \cellcolor{insec}\usebox\FilteringMemExample
		& \cellcolor{sec}\usebox\FilteringExample\\
		\cellcolor{header}\tabcenter{(A) } 
		& \cellcolor{header}\tabcenter{(B) } \\
		\hline
	\end{tabular}
	\caption{Counterexamples for Crypto-Erasure and Paralocks.}
	\label{fig:examples}
\end{figure}

For Gradual Release, it fails on the following secure program, where $ h, h1: \High
$ and $ l, l2: \Low $.
\begin{align*}
&\cod{if}~(h)~\cod{then}~ l2:= \cod{declassify}(h1); \\
& l:=0;
\end{align*}

This example might seem insecure on the surface, as the branch condition $h$
was not part of the $\cod{declassify}$ expression. But in the formal semantics
(Section~\ref{sec:gradualrelease}), a release event declassifies all information
in the program (i.e., Gradual Release does not provide a precise bound on the
released information \revise{as pointed out in }~\cite{askarov2007, 
Askarov:2009csf}). The program
is secure since $h1$ is assigned to $l2$ when both $h$ and $h1$ are
declassified by the release event.

To check if a similar issue also exists in Paralocks, whose security condition
is a generalization of the Gradual Release, we created a Paralock version of
the same code, as shown in Figure~\ref{fig:examples}-(B). Thanks to the cleaner
syntax of Paralocks, it is more obvious that the program is secure: $ h $ and
$h1 $ have the same lock set $ \{D\} $. Lock $ D $ is opened before the if
statement, allowing value of both $ h $ and $ h1 $ to flow to $ l, l2 $.  So
the assignment in the if branch is secure. After that, only a constant $ 0 $ is
assigned to $ l $ when the lock $D$ is closed. 
However, the Paralock implementation rejects this program as insecure. 

To understand why, Paralocks requires the knowledge of an attacker remains the
same if the current lock state is a subset of the lock set that the attacker
have.  We are interested in attacker $ A_1 = (a, \emptyset) $, who has an empty
lock set.  When lock $ D $ is open, since $ \{D\} \not \subseteq \emptyset $,
there is no restriction for the assignment $l2:=h1$. However, for the
assignment $l:=0$, the current lock set is $ \emptyset $, which is a subset
of $ A $'s lock set ($ \emptyset $). That is, for all the executions, the
attacker $ A $'s knowledge should not change by observing the output event
from assignment $ l:=0 $.
However, this does not hold for the execution starting with $ h=0 $.  The
initial knowledge of attacker $ A  $ knows nothing about $ h $ or $ h1 $ since
they are protected by lock $ D $. With $ h=0 $, the assignment in the branch is 
not
executed. The attacker only observes the output from $ l:=0 $. By observing that
output, the attacker immediately learns that $ h=0 $. Therefore, Paralock
rejected this program as insecure.

\section{Related Work}
The most related works are those present high-level discussions on what/how
end-to-end secure confidentiality should look like for some dynamic security
policy. The major ones are already discussed and compared in the paper. 

\revise{
To precisely describe a dynamic policy, {RIF}~\cite{kozyri2020, kozyri2019}
uses reclassification relation to associate label changes with proram outputs.
While this approach is highly expressive, writing down the correct relation
with regards to numerous possible outputs is arguably a time-consuming and
error-prone task. Similarly, flow-based declassification~\cite{rocha2010}
uses a graph to pin down the exact paths leading to a declassification.
However, the policy specification is tied up to the literal implementation of
a program, which might limit its use in practice.}

Bastys et al.~\cite{Bastys:2018plas} present six informal design principles
for security definitions and enforcements.  They summarize and categorize 
existing works to build a road map for the state-of-art. Then, from the top-down
view, they provide guidance on how to approach a new enforcement or definition.
In contrast, the framework and the benchmark proposed in this paper are
post-checks after one definition is formalized.

Recent work~\cite{Chudnov:2018csf} presents a unified framework for expressing
and understanding for downgrading policies. Similar to 
Section~\ref{sec:semantics}, the goal of the framework is to make obvious the
meaning of existing work. Based on that, they move further to sketch safety
semantics for enforcement mechanism. However, they do not provide a define a
formalization framework that allows us to compare various policies at their
semantics level.

Many existing work~\cite{McCall:2018csf, Gollamudi:2016oopsla, Buiras:2015plas}
reuses or extends the representative policies we discussed in this paper.  They
adopt the major definition for their specialized interest, which are irrelevant
to our interest.  Hunt and Sands~\cite{hunt2008} present an interesting
insight on  erasure, but their label and final security definition are attached
to scopes, which is not directly comparable with the end-to-end definitions
discussed in this work.  Contextual
noninterference~\cite{Polikarpova:2016arxiv} and 
\revise{{facets}~\cite{austin2012}} use 
dynamic labels to keep track of information flows in different branches.
The purpose of those labels is to boost flow- or path-sensitivity, not
intended for dynamic policies.

\section{Conclusion and Future Work}
\label{sec:future}
We present the first formalization framework that allows apple-to-apple
compassion between various dynamic policies. The comparison sheds light on
new insights on existing definitions, such as the distinguishing between
transient and persistent policies, as well as motivates \policyname{}, a new
general dynamic policy proposed in this work. Moreover, we built a new
benchmark for testing and understanding dynamic policies in general.

For future work, we plan to investigate semantic security condition of dynamic
information flow methods, especially those use dynamic security labels. Despite
the similarity that security levels are mutable, issues such as label channels
might be challenging to be incorporate in our formalization framework.
Moreover, \policyname{} offers a semantic definition for information-flow
security, but checking it on real programs is infeasible unless only small
number of traces are produced. We plan to develop a static type system to check
\policyname{} in a sound and scalable manner.

\revise{
Another future direction is to fully support partial release with
expression-level specification. 
However, doing so is tricky since the expressions might have conflicting
specifications.  For example, consider a specification $ x,y: \High $ and $
x+y, x-y:\Low $. It states that the values of $ x $ and $ y $ are secrets, but
the values of $ x+y $ and $ x-y $ are public. Mathematically, learning the
values of $ x+y $ and $ x-y $ can also reveal the concrete values of $ x $ and
$ y $. Thus, it becomes tricky to define security in the presense of
expression-level specification.  
}

\section{Acknowledgement}

We would like to thank the anonymous CSF reviewers for their constructive
feedback. This research was supported by NSF grants CNS-1942851 and
CNS-1816282.


\bibliographystyle{IEEEtranS}
\bibliography{main} 

\ifreport
\clearpage

\appendix

\subsection{Paralock in Table~\ref{tab:summary}}

Paralock~\cite{Broberg:2006esop,broberg2009,Broberg:2010popl} uses locks
to formalize the sensitivity of security objects. Paralock uses fine-grained
model to encode role-based access control systems.  Its covers both
declassification and revocation of information to a principal in the system.  As
described in Section~\ref{sec:paralock-label}, security specification is 
written as $ \{\lockset
\sat a;... \} $, where $ \lockset $ is a \emph{lock set} and $ a $ is an
\emph{actor}. An actor $ a $ is the base sensitivity entity of the model, which
is used to model a lattice level $ L $ in two-point lattice $ \{H, L\} $
in~\cite{broberg2009}, and a principal $ p $ in role-base access control
system in~\cite{Broberg:2010popl}. 

To formalize Paralock security, an attacker $ A = (a , \lockset) $ is modeled as
an actor $ a $ with a (static) set of open locks $ \lockset $.  To simplify
notation, we use $ \spec(x, a) = \lockset $ to denote the fact that $
\{\lockset \sat a;\}$ is part of the security policy of $ x $, otherwise $ 
\spec(x, a) = \top $. With respect to
an attacker $ A = (a , \lockset) $, a variable $ x $ is observable to $ A $ iff
$ \spec(x, a) \subseteq \lockset $, meaning that the attacker possesses more 
opened locks than what's required in the policy. 

To simplified the notations in this work, we extend the output event $ t $ to 
also record the current open locks. So, for a trace fragment $ \configs{c, 
m}\xrightarrow[]{b, v, \gamma} \configs{c', m'}$, 
it generates the output event $ t = \configs{b, v, \gamma, \Delta} $, where $ 
\Delta = \cod{unlock}({\configs{c,m}})$.

Let $\eset{A}$ be the set of
variables that are visible to $A$, and $\proj{\vec{t}}_A$ be the outputs that 
are visible to $ A  = (a_A, \lockset_{A})$:
\begin{align*}
\eset{A} ~\defn~ & \{ x \mid \forall x \in \Vars. ~\spec(x, a_A)  \subseteq 
\lockset_{A} \}\\
\proj{\vec{t}}_A ~\defn~ & \proj{\vec{t}}_{   \lambda {b, n,
\gamma, \Delta}.  \spec(b, a_A) \subseteq  \lockset_A  }
\end{align*}
Paralock security defines attacker's knowledge\footnote{We revised the 
definition for termination insensitivity. We note that Paralock 
also presents a different termination insensitive policy 
following ideas from\cite{Askarov:2008}. However, here we follow Gradual 
Release to define terminated insensitive knowledge by taking a 
intersection with the initial memory of traces that terminates.} as follows:
\begin{multline*}
k_{\cod{PL}}(c, m, \vec{t}, A) = \{m'~|~ m' \loweq_{\eset{A}} m \\
\AND \configs{c,m'} \xrightarrow{\vec{t_1}}{^*} \configs{c',m''} 
\xrightarrow{\vec{t_2}}{^*} m'''
\AND \proj{\vec{t_1}}_A = \proj{\vec{t}}_{A} \}
\end{multline*}

Paralock security semantics extends that of gradual release, by treating
``unlock'' events as releasing events:
\begin{Definition}[Paralock Security] 
\label{def:paralock-original}
\label{def:paralock}
A program $ c $ is Paralock secure if for 
any attacker $ A = \configs{a, \lockset} $, the attacker's knowledge
remains unchanged whenever $\cod{unlock}(\trace_{[i]}) \subseteq 
\lockset_A$:
\begin{multline*}
\forall c, m, m', \vec{t},t', a, \lockset_A, A, i. ~\\
\configs{c, m} \xrightarrow{\vec{t}} \configs{c', m'}  
\xrightarrow{t}\configs{c'', m''}
\AND A = \configs{a, \lockset_A}  
\AND \\\cod{unlock}(\configs{c'', m''}) \subseteq \lockset_A\\
\sat k_{\cod{PL}}(c, m, \vec{t}\cdot t',  A) 
= k_{\cod{PL}}(c, m, \vec{t}, A)
\end{multline*}
\end{Definition}
We use the memory closure on $ A $ for memory that looks the same to attacker $ 
A $:
\begin{multline*}
\closure{m}_{A} ~\defn~ \{ m'~|~\forall m'.~ \forall  x \in \Vars. \\
~\spec(x, a_A)  \subseteq \lockset_{A} \sat m(x) = m'(x) \}
\end{multline*}
The conversion of Definition~\ref{def:paralock} to our framework is
straightforward. 
\[ 
\sim_{\cod{PL}} \defn \{ (\vec{t_1}, \vec{t_2}) ~|~ \proj{\vec{t_1}}_{A}\text{ 
prefix of } \proj{\vec{t_2}}_{A} \} ~~~~ 
\equiv_{\cod{PL}} 
\defn = \]
\[ \allowence_{\cod{PL}} \defn 
\begin{cases}
\mathcal{K}(c, \vec{t}^{[:i-1]}, \sim_{\cod{PL}})
~\cap~ \closure{m}_{A }, & \vec{t}^{[\len{\vec{t}}]}.\Delta \subseteq 
\lockset_{A}\\
\closure{m}_\emptyset, &\text{otherwise}
\end{cases}
\]
\begin{Lemma}
\label{lemma:paralock-eq}
With $\sim \defn \sim_{\cod{PL}}$ and $\allowence \defn 
\allowence_{\cod{PL}}  $,
Definition~\ref{def:framework} is equivalent to 
Definition~\ref{def:paralock}.
\end{Lemma}
\subsection{Equivalence Proof for Table~\ref{tab:summary}}

We first introduce a useful lemma which allows us to rewrite cast the orginal
knowledge defintion in~\cite{askarov2007} to the knowledge defitnion
$\mathcal{K}$ in this paper.

\begin{Lemma}
\label{lem:convert}
Let $k$ be defined as in Equation~\ref{def:k_GR} and $\mathcal{K}$ be 
defined as
in Equation~\ref{eq:kout}, then we have
\begin{multline*}
\forall c, m, L, \Gamma, \vec{t}, M.~\\
M \subseteq \closure{m}_{L, \Gamma} 
\AND 
k(c,m,\vec{t},L,\Gamma)\subseteq
M \implies \\ k(c,m,\vec{t},L,\Gamma)=M \iff 
\mathcal{K}(c,\vec{t},\sim_{\cod{NI}})\supseteq M 
\end{multline*}
\end{Lemma}
\begin{proof}
By definition, we know
\[ k(c,m,\vec{t},L,\Gamma)=\mathcal{K}(c,\vec{t}, \sim_{\cod{NI}})\cap 
\closure{m}_{L, \Gamma} \]
\begin{itemize}
\item case $ \Longrightarrow $: we know
\begin{align*}
M =  k(c,m,\vec{t},L,\Gamma)  ~=~ & \mathcal{K}(c,\vec{t}, \sim_{\cod{NI}}) 
\cap 
\closure{m}_{L, \Gamma}\\
\mathcal{K}(c,\vec{t}, \sim_{\cod{NI}}) ~\supseteq~ & \mathcal{K}(c,\vec{t}, 
\sim_{\cod{NI}}) \cap 
\closure{m}_{L, \Gamma}
\end{align*}
Thus, we have $ \mathcal{K}(c,\vec{t}, \sim_{\cod{NI}}) ~\supseteq M  $. 

\item case $\Longleftarrow$: we know 
\begin{align*}
k(c,m,\vec{t},\Gamma)~=~
&\mathcal{K}(c,\vec{t}, \sim_{\cod{NI}})~\cap ~\closure{m}_{L,\Gamma} \\
M  ~\subseteq~ & \mathcal{K}(c,\vec{t}, \sim_{\cod{NI}})  \\
M ~\subseteq~ &\closure{m}_{L, \Gamma} \\
M \cap M  ~\subseteq~ &\mathcal{K}(c,\vec{t}, \sim_{\cod{NI}})~\cap 
~\closure{m}_{L,\Gamma} 
\end{align*}
Thus, we know $ k(c,m,\vec{t},L, \Gamma) \supseteq M $. From assumption $ 
k(c,m,\vec{t},\Gamma) \subseteq M  $, we know $ k(c,m,\vec{t},L,\Gamma) = M $.
\end{itemize}	
%
So, we have  $ k(c,m,\vec{t},L, \Gamma)=M \iff 
\mathcal{K}(c,\vec{t}, \sim_{\cod{NI}})\supseteq M  $.
\end{proof}
\subsubsection{Gradual Release}
\noindent\textbf{\textit{Lemma~\ref{lemma:gradualrelease-eq}.}} 
With $\sim \defn \sim_{\cod{GR}}$ and $\allowence \defn \allowence_{\cod{GR}}$,
Definition~\ref{def:framework} is equivalent to
Definition~\ref{def:gradualrelease}:
\begin{multline*}
\forall c, m, L, i, \vec{t}.~ \configs{c,m} \termout \vec{t} \implies \\
\text{i not release event} \sat k(c,m,\vec{t}_{ 
	[:i]},L, \Gamma)=k(c,m,\vec{t}^{[:i-1]},L,\Gamma) \\ 
\iff  
\mathcal{K}(c,\vec{t}^{[:i]}, \sim_{\cod{GR}}) 
\supseteq \closure{m}_{L, ~\vec{t}^{[i]}.\gamma} 
\cap \mathcal{K}(c,\vec{t}_{ [:i-1]}, \sim_{\cod{GR}})
\end{multline*}
\begin{proof} From the encoding of Gradual Release, we know:
\begin{equation*}
\vec{t}^{[i]}.\gamma = 
\begin{cases}
\gamma_{\bot}, & \mbox{i is a release event} \\
\Gamma, &\mbox{i not a release event}
\end{cases}
\end{equation*}
\begin{itemize}
\item case when $ i $ is a release event: $ \vec{t}^{[i]}.\gamma 
= \gamma_{\bot} $. From the definition, we know $ \closure{m}_{L, 
	~\gamma_{\bot}}$ returns the singleton set $ \{m\} $.  \\
From $ \configs{c,m} \termout \vec{t} $ and the definition of $ \mathcal{K} $,  
we 
know $ \forall j.~m \in \mathcal{K}(c,\vec{t}^{[:j]}, \sim_{\cod{GR}}) $:
\begin{align*}
m ~\in~ &\mathcal{K}(c,\vec{t}^{[:i]}, \sim_{\cod{GR}})\\
m ~\in~ &\mathcal{K}(c,\vec{t}^{[:i-1]}, \sim_{\cod{GR}})\\
\{m\} ~=~ &\closure{m}_{L, ~\vec{t}^{[i]}.\gamma }
\end{align*}
Thus, both Definition~\ref{def:framework} and \ref{def:gradualrelease} 
are trivially true. 
\item case when $ i $ is not a release event: $ \vec{t}^{[i]}.\gamma 
= \Gamma $. From the definitions, we know $ \sim_{\cod{GR}} = \sim_{\cod{NI}} 
$. We know from the monotonicity of the knowledge that:
\begin{align*}
k(c,m,\vec{t}^{[:{i-1}]},L,\Gamma) ~\subseteq~& \closure{m}_{L, \Gamma }\\  
k(c,m,\vec{t}^{[:{i}]},L,\Gamma) ~\subseteq~& k(c,m,\vec{t}_{ 
	[:{i-1}]},L, \Gamma)
\end{align*}
So, we can instantiate Lemma~\ref{lem:convert} with:
\[ M: = k(c,m,\vec{t}^{[:{i-1}]},L,\Gamma), ~~~~\vec{t} := \vec{t}^{[:i]}\]
and we get:
\begin{multline*}
k(c,m,\vec{t}^{[:{i}},L,\Gamma) =  k(c,m,\vec{t}^{[:i-1]},L,\Gamma)\\
\iff~~ \mathcal{K}(c,\vec{t}^{[:i]}, \sim_{\cod{GR}}) \supseteq 
k(c,m,\vec{t}^{[:{i-1}]},L,\Gamma)
\end{multline*}
By definition, we know 
\[ k(c,m,\vec{t}^{[:i-1]},L,\Gamma)=
\closure{m}_{L, \Gamma} \cap  \mathcal{K}(c,\vec{t}^{[:i-1]}, 
\sim_{\cod{GR}}) \]
Thus, when $ i $ is not a release event, we have:
\begin{multline*}
k(c,m,\vec{t}^{[:i]},L,\Gamma)=k(c,m,\vec{t}^{[:i-1]},L,\Gamma) \\ 
\iff  
\mathcal{K}(c,\vec{t}^{[:i]}, \sim_{\cod{GR}}) 
\supseteq \closure{m}_{L, ~\vec{t}^{[i]}.\gamma} 
\cap \mathcal{K}(c,\vec{t}^{[:i-1]}, \sim_{\cod{GR}})
\end{multline*}
\end{itemize}
Therefore, Definition~\ref{def:framework} is equivalent to 
Definition~ \ref{def:gradualrelease}. 
\end{proof}

\subsubsection{Tight Gradual Release} 

\noindent\textbf{\textit{Lemma~\ref{lemma:tightGR-eq}.}} 
With $\sim \defn \sim_{\cod{TGR}}$, $\equiv\defn \equiv_{\cod{TGR}}$ and
$\allowence \defn \allowence_{\cod{TGR}}$, Definition~\ref{def:framework} is 
equivalent to Definition~\ref{def:tightGR}.
\begin{multline*}
\forall i. ~1 \leq i \leq \len{\vec{t}}.~  \\
(\closure{m}_{L, \Gamma} \cap \closure{m}_{E_i} )
\subseteq k(c,m, \vec{t}^{[:i]}, L, \Gamma)\\
\iff \\
\mathcal{K}(c,\vec{t'}^{[:i]}, \sim_{\cod{TGR}}) \supseteq  \closure{m}_{L, 
\vec{t}^{[i]}.\gamma}
\end{multline*}

\begin{proof} The encoding limited $ E_i $ to a variable set $ X_i $, thus, we 
assumes $ E_i = X_i $.
From the definition, we know that
\begin{equation}
\label{eq:tight-eq-1}
k(c,m,\vec{t}, L, \Gamma) = \mathcal{K}(c,\vec{t'}^{[:i]}, \sim_{\cod{TGR}}) 
\cap \closure{m}_{L, \Gamma}
\end{equation}
From encoding, we know that 
\begin{equation}
\label{eq:tight-eq-2}
\closure{m}_{L, \vec{t}^{[i]}.\gamma} = (\closure{m}_{L, \Gamma} \cap 
\closure{m}_{E_i} )
\end{equation}
\begin{itemize}
\item  case $ \Longrightarrow $: From Equation~\ref{eq:tight-eq-2} and the 
assumption, we know 
\[
\closure{m}_{L, \vec{t}^{[i]}.\gamma} \subseteq  k(c,m, \vec{t}^{[:i]}, L, 
\Gamma)
\]
From Equation~\ref{eq:tight-eq-1}, we know
\[
 k(c,m, \vec{t}^{[:i]}, L, \Gamma) \subseteq  \mathcal{K}(c,\vec{t'}^{[:i]}, 
 \sim_{\cod{TGR}})
\]
Therefore, we have $  \closure{m}_{L, \vec{t}^{[i]}.\gamma} \subseteq 
\mathcal{K}(c,\vec{t'}^{[:i]}, 
\sim_{\cod{TGR}})  $.
\item  case $ \Longleftarrow $: By taking an intersection with $ 
\closure{m}_{L, \Gamma} $ on both side of the assumption, we have:
\[
\mathcal{K}(c,\vec{t'}^{[:i]}, \sim_{\cod{TGR}}) \cap \closure{m}_{L, \Gamma}  
\supseteq  \closure{m}_{L,\vec{t}^{[i]}.\gamma} \cap \closure{m}_{L, \Gamma} 
\]
Apply Equation~\ref{eq:tight-eq-1} to the left and Equation~\ref{eq:tight-eq-2} 
to the right:
\begin{align*}
k(c,m, \vec{t}^{[:i]}, L, \Gamma) \supseteq~ (\closure{m}_{L, \Gamma} &\cap 
\closure{m}_{E_i} ) \cap \closure{m}_{L, \Gamma}  \\
=&  ~ \closure{m}_{L, \Gamma} \cap \closure{m}_{E_i}  
\end{align*}
Thus, we have $ k(c,m, \vec{t}^{[:i]}, L, \Gamma) \supseteq 
(\closure{m}_{L, \Gamma} \cap \closure{m}_{E_i}) $.
\end{itemize} 
Therefore, Definition~\ref{def:framework} is 
equivalent to Definition~\ref{def:tightGR}.
\end{proof}

\subsubsection{According to Policy p}
\noindent\textbf{\textit{Lemma~\ref{lemma:according-eq}.}} 
With $\sim \defn \sim_{\cod{AP}}$, $\allowence \defn
\allowence_{\cod{AP}}$, and outside equivalence $ \equiv \defn  
\equiv'_{\cod{AP}}$, Definition~\ref{def:framework} is equivalent to
Definition~\ref{def:niaccord}.
\begin{proof}
	First, we convert a security levels $ L $ from the Denning's 
	style to our attacker levels $ l $ as described in 
	Section~\ref{sec:syntax},  
	and outputs any intermediate memory of the trace to its variable's level. 
	That 
	is,
	\begin{multline*}
	\forall c, m_0, m', c_i, m_i, i, \sentity, \Gamma. \\
	\trace = \configs{c, m_0}_{\gamma_0} \rightarrow^* 
	\configs{c_i, m_i}_{\gamma_i}  \rightarrow^* m' \AND \trace_{[i]} = 
	\configs{c_i, m_i}_{\gamma_i}  \\
	\iff \\
	\configs{c, m_0} \termout \vec{t}  \\
	\AND \vec{t}^{[i]} = \{ \configs{ch, n, \gamma}~|~ ch = \sentity 
	\AND n = m_i(\sentity) \AND \gamma = \gamma_i  \} 
	\end{multline*}
	We note that our normal $ \vec{t}^{[i]} $ returns a single output event, 
	say 
	some $ t =\configs{ch, n, \gamma} $. But here we overload $ \vec{t}^{[i]} $ 
	to 
	return a set of output events that output all values on memory $ 
	\trace_{[i]} $.  
	Thus, with all values on memory outputted, we have:
	\begin{multline*}
	\forall c, m_1, m_2, m_1', m_2', \trace, \trace', \vec{t_1}, \vec{t_2}. \\
	\AND \trace =  \configs{c, m_1} \xrightarrow{} m_1'
	\AND \trace' = \configs{c, m_2} \xrightarrow{} m_2' \\
	\AND \configs{c, m_1} \termout \vec{t_1}
	\AND \configs{c, m_2} \termout \vec{t_2} \implies \\
	{\trace}_{[i]} \loweq_{l} \trace'_{[j]} 
	\iff \proj{\vec{t}^{[i]}}_l = \proj{\vec{t}^{[j]}}_l. 
	\end{multline*} 
	Thus, we rewrite Definition~\ref{def:niaccord} in following two-run style:
	\begin{multline*}
	\forall c, m_1, m_2, l, p, \vec{t_1}, \vec{t_2}.\\
	m_2 \in  \closure{m_1}_{p} 
	\AND \configs{c, m_1}\termout \vec{t}_1   
	\AND \configs{c, m_2} \termout \vec{t}_2 \implies \\
	\exists R. \Big(
	\forall (i,j)\in R. ~ \vec{t_1}_{[i]} \in  \proj{\vec{t_1}}^{p,l} 
	\AND \vec{t_2}_{[j]} \in  \proj{\vec{t_2}}^{p,l} \\
	\sat \proj{\vec{t}_1}_{l} = \proj{\vec{t}_2}_{l} 
	\Big)
	\end{multline*}
	We combine the two filters and assume $ R' $ as $ R $ after filtering:
	\begin{multline*}
	\forall c, m_1, m_2, l, p, \vec{t_1}, \vec{t_2}.\\
	m_2 \in  \closure{m_1}_{p} 
	\AND \configs{c, m_1}\termout \vec{t}_1   
	\AND \configs{c, m_2} \termout \vec{t}_2 \implies \\
	\exists R'. \Big( 
	\forall (i,j)\in R'. 
	(\proj{\vec{t}_1}_{p,l})_{[i]} = (\proj{\vec{t}_2}_{p,l})_{[j]}
	\Big)
	\end{multline*}
	With $ \mathcal{K}(c, \vec{t}, \sim_{\cod{AP}}) $ unfolded as below:
	\begin{multline*}
	\mathcal{K}(c, \vec{t}, \sim_{\cod{AP}}) = \{ m_2 ~|~ \forall m_2, 
	\vec{t_2}.~
	\configs{c, m_2} \termout \vec{t_2} \\
	\AND \exists R'. \forall (i,j)\in R'.~(\proj{\vec{t}}_{p,l})_{[i]} 
	= (\proj{\vec{t}_2}_{p,l})_{[j]}  \}
	\end{multline*}
	We can further rewrite the definition as follow:
	\begin{multline*}
	\forall c, m_1, m_2, l, p, \vec{t_1}.\\
	m_2 \in  \closure{m_1}_{p} 
	\AND \configs{c, m_1}\termout \vec{t}_1 \implies 
	m_2 \in \mathcal{K}(c, \vec{t_1}, \sim_{\cod{AP}})
	\end{multline*}
	That is,
	\begin{multline*}
	\forall c, m_1, l, p, \vec{t_1}. \configs{c, m_1}\termout \vec{t}_1 \AND   
	\closure{m_1}_{p} \subseteq    
	\mathcal{K}(c, \vec{t_1}, \sim_{\cod{AP}})
	\end{multline*} 
	We note that only the equivalence relation in $ \sim_{\cod{AP}} $ is $ 
	\equiv_{\cod{AP}} $. The equivalence relation in $ \vec{t'} \equiv \vec{t}$ 
	in 
	Definition~\ref{def:framework} in this case is not $ \equiv_{\cod{AP}} $, 
	but $ 
	\equiv'_{\cod{AP}} \defn  \{ (\vec{t_1}, \vec{t_2}) ~|~\vec{t_1} = 
	\vec{t_2} \} $.
\end{proof}

\subsubsection{Cryptographic Erasure}
\noindent\textbf{\textit{Lemma~\ref{lemma:crypto-eq}.}} 
	With $\sim \defn \sim_{\cod{CE}}$, $\equiv\defn \equiv_{\cod{CE}}$ and
	$\allowence \defn \allowence_{\cod{CE}}$, Definition~\ref{def:framework} 
	with
	adjusted attack model is equivalent to Definition~\ref{def:crypto}.

\begin{multline*}
\forall c, m, \gamma_0, c_i, m_i, \gamma_i, c_n, m_n, \gamma_n, m',\vec{t_1}, 
\vec{t_2}, l, i, j, n.~\\
\configs{c,m}_{\gamma_0} \xrightarrow{\vec{t_1}}
\configs{c_i,m_i}_{\gamma_i} \xrightarrow{\vec{t_2}}
\configs{c_n,m_n}_{\gamma_n} \rightarrow^* m' \implies\\
k_{\cod{CE}}(c, L, \sim_{\cod{CE}}) \supseteq \bigcap_{i\leq j \leq n} 
\closure{m}_{L, \gamma_j}  
\iff \\
\mathcal{K}(c, \vec{t_2}, \sim_{\cod{CE}}) \supseteq \bigcap_{t_n \in 
\vec{t_2}} 
\closure{m}_{L, t_n.\gamma}
\end{multline*}
\begin{proof}
We note that in Definition~\ref{def:crypto}, $ \gamma_j $ are state policies 
attached to the configurations, not from the output event $ \vec{t} $. 
According to the definition, $ \vec{t} $ does not contain empty events. In 
Definition~\ref{def:crypto},  it takes $ n-j $ steps to generate output 
sequence $ \vec{t_2} $, we know $ n-j\geq \len{\vec{t_2}} $. 
We first show that the right hand side allowance defined using $ \gamma_j $ is 
the same as using state policy from the output sequence $ \vec{t} $:
\begin{multline}
\label{eq:crypto-mem-out}
k_{\cod{CE}}(c, L, \vec{t_2}) \supseteq 
\bigcap_{i\leq j \leq n} \closure{m}_{L, \gamma_j} 
\\ \iff  
k_{\cod{CE}}(c, L, \vec{t_2}) \supseteq 
\bigcap_{t_n \in \vec{t_2}} \closure{m}_{L, t_n.\gamma}
\end{multline}
\begin{itemize}
\item case $ \implies $:
Crypto~\cite{askarov2015} supports only erasure 
policy (and static policy). That is, the sensitivity of any security entity is 
monotonically increasing: 
\[
\forall j \in [i, n]. \gamma_{j} \pref \gamma_{j+1}
\]
From the definition of memory closure, we know:
\[
\forall \gamma_1, \gamma_2.~\gamma_{1} \pref \gamma_{2} \implies 
\closure{m}_{L, \gamma_1} \subseteq  \closure{m}_{L, \gamma_2}
\]
Thus, we know:
\begin{align*}
\bigcap_{i\leq j \leq n} \closure{m}_{L, \gamma_j} 
&~=~\closure{m}_{L, \gamma_i}\\
\bigcap_{t_n \in \vec{t_2}} \closure{m}_{L, t_n.\gamma} 
&~=~\closure{m}_{L, \vec{t_2}_{[0]}.\gamma}
\end{align*}
From the definitions, we know $ \vec{t_2}_{[0]}.\gamma = \gamma_i $ if $ 
\configs{c_i,m_i}_{\gamma_i} $ does not immediately generates an empty 
output event. Otherwise, if the first non-empty event is generated at 
configuration $ \configs{c_{i'}, m_{i'}}_{\gamma_{i'}}, (i<i'<n) $, we know:
\[\closure{m}_{L, \gamma_i} \subseteq \closure{m}_{L, \gamma_{i'}} = 
\closure{m}_{L, \vec{t_2}_{[0]}.\gamma} \]
We can instantiate Definition~\ref{def:crypto} for $ i := i' $, and we get:
\[
k_{\cod{CE}}(c, L, \vec{t_2}) \supseteq 
\bigcap_{i'\leq j \leq n} \closure{m}_{L, \gamma_{j}}  = \closure{m}_{L, 
\gamma_{i'}} = 
\closure{m}_{L, \vec{t_2}_{[0]}.\gamma}
\]
Thus, we have $ k_{\cod{CE}}(c, L, \vec{t_2}) \supseteq 
\bigcap_{t_n \in \vec{t_2}} \closure{m}_{L, t_n.\gamma} $. 
\item case $ \impliedby $: from $  n-j\geq \len{\vec{t_2}} $, we know:
\begin{align*}
\forall t_n \in \vec{t_2}.~ \exists j' \in [i,n].~ \gamma_{j'} &~=~ t_n.\gamma
\\
\{t_n.\gamma ~|~t_n \in \vec{t_2}  \} &~\subseteq~ \{ \gamma_j ~|~i\leq j \leq
n   \}\\
\bigcap_{t_n \in \vec{t_2}}
\closure{m}_{L, t_n.\gamma} &~\supseteq~ \bigcap_{i\leq j \leq n}
\closure{m}_{L, \gamma_j}
\end{align*}
Thus, we have $ k_{\cod{CE}}(c, L, \vec{t_2}) \supseteq 
\bigcap_{i\leq j \leq n} \closure{m}_{L, \gamma_j}  $.
\end{itemize}
Therefore, we know Equation~\ref{eq:crypto-mem-out} is true. \\
Now we convert a security level $ L $ from Denning's style to our attacker 
level $ l $ as described in Section\ref{sec:syntax}. 
Let $ \closure{c} \defn \{m ~|~ \exists \vec{t}.~ \configs{c,m} \termout 
\vec{t}  \} $ denote the set of memory that terminates. From definition we know:
\[
\mathcal{K}(c, \vec{t}, \sim_{\cod{CE}})  = k_{\cod{CE}}(c, L, \vec{t}) \cap 
\closure{c}  
\] 
For the interest of a termination-insensitive policy, we can ignore the 
difference made by the terminated set $ \closure{c} $. Thus, we assumes $ 
\mathcal{K}(c, 
\vec{t}, \sim_{\cod{CE}}) = k_{\cod{CE}}(c, L, \vec{t})  $.
\end{proof}

\subsubsection{Forgetful Attacker}
\noindent\textbf{\textit{Lemma~\ref{lemma:forgetful-eq}.}} 
With $\sim \defn \sim_{\cod{FA}}$ and $\allowence \defn 
\allowence_{\cod{FA}}  $,
Definition~\ref{def:framework} is equivalent to 
Definition~\ref{def:forgetful}.

\begin{proof}
	In the forgetful attacker\cite{askarov2012}, the sensitivity level is 
	changed by  $ \cod{setPolicy} $ command.  
	Recall from our encoding, $ \cod{setPolicy} $
	is encoded using security commands and generates a security event, but no 
	output event. So, there is no sensitivity change between the two states 
	that 
	generates an output. That is, for the output event $ t' $ in the trace:
	\[
	\configs{c, m}\xrightarrow{\vec{t} \cdot t'}{}^* 
	\configs{c',m'} \xrightarrow{\configs{b,v, \gamma'}\cdot ...}{}^* \sat 
	\]
	 We know $ t'.\gamma =  \gamma' $ and therefore, we have \[  
	 \closure{m}_{L, \gamma'} = \closure{m}_{L, ~ t'.\gamma} \] 
	Definition~\ref{def:forgetful} is rephrased as:
	\begin{multline*}
	\forall c, m, L, i, \vec{t}. \configs{c,m} \termout \vec{t} \implies\\
	k_{\cod{FA}}(c, L, \cod{Atk}, \vec{t}^{[:i]}) 
	\subseteq 
	k_{\cod{FA}}(c, L, \cod{Atk}, \vec{t}^{[:i-1]}) 
	\cap \closure{m}_{L, ~ \vec{t}^{[i]}.\gamma}
	\end{multline*}
	By definition, we know:
	\[
	k_{\cod{FA}}(c, L , \cod{Atk}, \vec{t}) = \mathcal{K}(c, \vec{t}, 
	\sim_{\cod{FA}})
	\]
	Thus, we know Definition~\ref{def:forgetful} is equivalent to 
	Definition~\ref{def:framework}.
\end{proof}


\subsubsection{Paralock}
\noindent\textbf{\textit{Lemma~\ref{lemma:paralock-eq}.}} 
With $\sim \defn \sim_{\cod{PL}}$ and $\allowence \defn 
\allowence_{\cod{PL}}  $,
Definition~\ref{def:framework} is equivalent to 
Definition~\ref{def:paralock}.
\begin{multline*}
\forall c,m, \vec{t}, \vec{t'}, i, A.~~
\configs{c, m} \termout \vec{t} \AND \vec{t'} = \vec{t}^{[:i]} \implies \\
\vec{t}^{[i]}.\Delta \subseteq \lockset_{A} \sat\\ 
k_{\cod{PL}}(c,m,\vec{t}^{[:i]}, 
A) = k_{\cod{PL}}(c,m,\vec{t}^{[:i-1]}, A)\\
\iff \\
\mathcal{K}(c,\vec{t}^{[:i]}, \sim_{\cod{PL}}) 
\supseteq \mathcal{K}(c,\vec{t}^{[:i-1]}, \sim_{\cod{PL}}) \cap \closure{m}_{A}
\end{multline*}
\begin{proof}
We omit the case when $ \vec{t}^{[i]}.\Delta \not \subseteq \lockset_{A} $ 
since 
both definitions are trivially true. By Definition, we know:
\begin{equation*}
\label{eq:connect}
\forall j.~k_{\cod{PL}}(c, m, \vec{t}^{[:j]}, A) = 
\mathcal{K}(c, \vec{t}^{[:j]}, \sim_{\cod{PL}}) \cap \closure{m}_A
\end{equation*}
\begin{itemize}
\item  case $ \implies $: we know:
\begin{align*}
k_{\cod{PL}}(c, m, \vec{t}^{[:i-1]}, A) ~=~
& k_{\cod{PL}}(c, m, \vec{t}^{[:i]}, A) \\
k_{\cod{PL}}(c, m, \vec{t}^{[:i]}, A) ~=~
& \mathcal{K}(c, \vec{t}^{[:i]}, \sim_{\cod{PL}}) \cap \closure{m}_A \\
\mathcal{K}(c, \vec{t}^{[:i]}, \sim_{\cod{PL}})  ~\supseteq~ 
&\mathcal{K}(c, \vec{t}^{[:i]}, \sim_{\cod{PL}}) \cap \closure{m}_A
\end{align*}
Thus, we have \[ \mathcal{K}(c, \vec{t}^{[:i]}, \sim_{\cod{PL}})  \supseteq 
k_{\cod{PL}}(c, m, \vec{t}^{[:i-1]}, A)\]
With $ k_{\cod{PL}}(c, m, \vec{t}^{[:i-1]}, A) = 
\mathcal{K}(c, \vec{t}^{[:i-1]}, \sim_{\cod{PL}}) \cap \closure{m}_A $, we get 
$ \mathcal{K}(c,\vec{t}^{[:i]}, \sim_{\cod{PL}}) 
\supseteq \mathcal{K}(c,\vec{t}^{[:i-1]}, \sim_{\cod{PL}}) \cap \closure{m}_{A} 
$.
\item  case $ \impliedby $: we know:
\begin{align*}
\mathcal{K}(c,\vec{t}^{[:i]}, \sim_{\cod{PL}}) 
~\supseteq~& \mathcal{K}(c,\vec{t}^{[:i-1]}, \sim_{\cod{PL}}) \cap 
\closure{m}_{A}\\
k_{\cod{PL}}(c, m, \vec{t}^{[:i-1]}, A) ~=~& 
\mathcal{K}(c, \vec{t}^{[:i-1]}, \sim_{\cod{PL}}) \cap \closure{m}_A
\end{align*}
Thus, we have 
\begin{align}
\label{eq:para1}
\mathcal{K}(c,\vec{t}^{[:i]}, \sim_{\cod{PL}}) \supseteq 
k_{\cod{PL}}(c, m, \vec{t}^{[:i-1]}, A)
\end{align}

We know $ \closure{m}_A $ is the initial knowledge of $ A $ before observing 
any 
output event. From the monotonicity of Paralock knowledge, we know:
\begin{align}
\label{eq:para2}
\closure{m}_A \supseteq k_{\cod{PL}}(c, m, \vec{t}^{[:i-1]}, A)
\end{align}
By taking an intersection on both side of Equation~(\ref{eq:para1}) and 
Equation~(\ref{eq:para2}), we have:
\begin{align*}
&(\mathcal{K}(c,\vec{t}^{[:i]}, \sim_{\cod{PL}})  \cap \closure{m}_A)  \\
\supseteq~
&(k_{\cod{PL}}(c, m, \vec{t}^{[:i-1]}, A) \cap k_{\cod{PL}}(c, m, 
\vec{t}^{[:i-1]}, A))\\
=~
& k_{\cod{PL}}(c, m, \vec{t}^{[:i-1]}, A)
\end{align*}
Thus, we have 
\[
k_{\cod{PL}}(c, m, \vec{t}^{[:i-1]}, A) \subseteq \mathcal{K}(c,\vec{t}^{[:i]}, 
\sim_{\cod{PL}})  \cap \closure{m}_A 
\]
By Definitions, we have:
\begin{align*}
k_{\cod{PL}}(c, m, \vec{t}^{[:i]}, A) ~=~
& \mathcal{K}(c, \vec{t}^{[:i]}, \sim_{\cod{PL}}) \cap \closure{m}_A
\end{align*}
Thus, we know:
\begin{align*}
k_{\cod{PL}}(c, m, \vec{t}^{[:i-1]}, A) ~\subseteq~ k_{\cod{PL}}(c, m, 
\vec{t}^{[:i]}, A) 
\end{align*}
From the monotonicity of the Paralock knowledge, we know 
\begin{align*}
k_{\cod{PL}}(c, m, \vec{t}^{[:i-1]}, A) ~\supseteq~ k_{\cod{PL}}(c, m, 
\vec{t}^{[:i]}, A) 
\end{align*}
Thus, we have:
\[
k_{\cod{PL}}(c, m, \vec{t}^{[:i-1]}, A) ~=~ k_{\cod{PL}}(c, m, 
\vec{t}^{[:i]}, A)
\]
\end{itemize}
Therefore, Definition~\ref{def:framework} is equivalent to 
Definition~\ref{def:paralock}.
\end{proof}

\fi
\end{document}